\documentclass[12pt]{article}
\usepackage{amssymb,amsmath,epsfig}
\begin{document}

\title{\bf Dynamical Instability of Gaseous Sphere in the Reissner-Nordstr\"{o}m Limit}
\author{M. Sharif \thanks {msharif.math@pu.edu.pk} and Saadia Mumtaz
\thanks{sadiamumtaz17@gmail.com}\\
Department of Mathematics, University of the Punjab,\\
Quaid-e-Azam Campus, Lahore-54590, Pakistan.}

\date{}
\maketitle

\begin{abstract}
In this paper, we study the dynamical instability of gaseous sphere
under radial oscillations approaching the Reissner-Nordstr\"{o}m
limit. For this purpose, we derive linearized perturbed equation of
motion following the Eulerian and Lagrangian approaches. We
formulate perturbed pressure in terms of adiabatic index by
employing the conservation of baryon numbers. A variational
principle is established to evaluate characteristic frequencies of
oscillations which lead to the criteria for dynamical stability. The
dynamical instability of homogeneous sphere as well as relativistic
polytropes with different values of charge in Newtonian and
post-Newtonian regimes is explored. We also find their radii of
instability in terms of the Reissner-Nordst\"{o}rm radius. We
conclude that dynamical instability occurs if the gaseous sphere
contracts to the Reissner-Nordst\"{o}rm radius for different values
of charge.
\end{abstract}

\section{Introduction}

It is a well-known fact that any relativistic model will be
physically interesting if it is stable under fluctuations. The
stability/instability of celestial objects has significant
importance in general relativity (GR). This study is closely related
to the evolution and structure formation of self-gravitating
objects. Initially, any stable gaseous mass remains in state of
hydrostatic equilibrium for which the gravitational force is counter
balanced by the internal pressure of the body acting in the opposite
direction. The effect of gravity over the internal pressure causes
the matter to collapse and the star contracts to a point under its
own gravitational force forming compact stars.

The dynamics of massive stars can be discussed in weak as well as
strong-field regimes. The idea of weak-field approximation
(Newtonian and post-Newtonian approximations (pN)) \cite{a} has
remarkable importance in the context of relativistic hydrodynamics.
The analysis of dynamical instability in strong-field regimes
becomes complicated due to non-linear terms, so the weak-field
approximation schemes are used as an effective tool. Chandrasekhar
\cite{b} was the pioneer who studied the dynamical instability of
Newtonian perfect fluid sphere approaching the Schwarzschild limit
in terms of adiabatic index. He used Eulerian approach for
hydrodynamic equations and developed a variational principle to find
characteristic frequencies applicable to the radial oscillations at
Newtonian and pN limits. He concluded that the system would be
dynamically stable or unstable according to the numerical value of
adiabatic index, i.e., $\Gamma>\frac{4}{3}$ or $\Gamma<\frac{4}{3}$,
respectively. The same author \cite{c} also investigated the
stability of gaseous sphere under radial and non-radial oscillations
at pN limit.

The dynamical instability of self-gravitating spherical objects has
been studied by using various techniques. Herrera et al. \cite{d}
explored dynamical instability of spherical collapsing system for
non-adiabatic fluid using perturbation scheme. They showed that heat
conduction increases the instability range in Newtonian limit but
decreases in pN limit. Later, many researchers \cite{e} discussed
the role of various physical factors on the dynamical instability of
spherical systems using perturbation scheme and found interesting
results.

The stability of self-gravitating objects in the presence of
electromagnetic field has a primordial history starting with
Rosseland \cite{1}. There is a general consensus that astrophysical
objects do not have charge in large amount \cite{2} but there are
some mechanisms which induce large amount of electric charge in
collapsing stars. Stettner \cite{3} showed that presence of net
surface charge enhances the stability of sphere with uniform
density. Glazer \cite{4} investigated the dynamical stability of
perfect fluid sphere pulsating radially with electric charge. Ghezzi
\cite{4a} studied stability of neutron stars and found that the
stars having a charge greater than the extreme value would explode.
Sharif and collaborators \cite{5} discussed the role of electric
charge in dynamical instability at Newtonian and pN regimes.

Polytropes are useful self-gravitating objects as they provide
simplified models for internal structures of stellar objects. The
polytropic equation of state deals with various fundamental
astrophysical issues \cite{6}. Tooper \cite{7} studied the internal
structure of gaseous sphere obeying polytropic equation of state and
obtained Newtonian polytropes using numerical solution of the
Lane-Emden equation. The effect of electromagnetic field on the
dynamics of polytropic compact stars has also been studied \cite{8}.
Herrera and Barreto \cite{9} analyzed both Newtonian as well as
relativistic polytropes in spherical symmetry. Recently, Breysse et
al. \cite{10} have discussed the dynamical instability of
cylindrical polytropic fluid systems under radial and non-radial
modes of oscillations.

In this paper, we study the dynamical instability of spherically
symmetric gaseous systems following Chandrasekhar's approach
\cite{b} in the vicinity of electromagnetic field. The paper is
organized as follows. The next section deals with matter
distribution and the Einstein-Maxwell field equations. In section
\textbf{3}, we discuss motion of the system under radial
oscillations following the Eulerian approach. Section \textbf{4}
provides the formulation of perturbed pressure and adiabatic index
in terms of Lagrangian displacement using conservation of baryon
number. In section \textbf{5}, we develop conditions for dynamical
instability of homogeneous sphere and relativistic polytropes.
Finally, we conclude our results in the last section.

\section{Field Equations and Matter Configuration}

We consider a spherically symmetric system in the interior region
given by
\begin{equation}\label{5}
ds^2=-e^{\nu}dt^2+e^{\lambda}dr^2+r^2(d\theta^2+\sin^2\theta
d\phi^2),
\end{equation}
where $\nu=\nu(t,r)$ and $\lambda=\lambda(t,r)$ are the
gravitational potentials. The corresponding Einstein field equations
can be written as
\begin{eqnarray}\label{6}
-\frac{8\pi
G}{c^4}T_{0}^0&=&\frac{1}{r^2}-\frac{1}{r^2}\frac{\partial}{\partial
r}(re^{-\lambda}),\\\label{7} -\frac{8\pi
G}{c^4}T_{1}^1&=&\frac{1}{r^2}-e^{-\lambda}(\frac{1}{r^2}+
\frac{1}{r}\frac{\partial \nu}{\partial r}),\\\label{8} \frac{8\pi
G}{c^4}T_{0}^1&=&\dot{\lambda}\frac{e^{-\lambda}}{r},
\end{eqnarray}
where dot denotes derivative w.r.t $t$. We assume the
energy-momentum tensor corresponding to charged perfect fluid in the
form
\begin{equation}\label{9}
T_{j}^i=(\sigma+p)u^iu_{j}+p\delta_{j}^i+\frac{1}{4\pi}
[F_{jk}F^{ik}-\frac{1}{4}\delta_{j}^iF_{kl}F^{kl}],
\end{equation}
where $u^i=\frac{dx^i}{ds}$ is the four-velocity, $p$ is the
pressure and $\sigma$ is the energy density. The electromagnetic
field tensor $F_{ij}$ can be defined in terms of four potential,
$F_{ij}=\Phi_{j;i}-\Phi_{i;j}$, which satisfies the Maxwell field
equations as
\begin{equation}\nonumber
F^{ij}_{~;j}=4\pi J^i, \quad F_{[ij,k]}=0,
\end{equation}
where $J^i=\rho u^i$ is the four current. The only non-vanishing
radial component of electromagnetic field tensor ($F^{01}=-F^{10}$)
implies that
\begin{equation}\nonumber
\frac{d\left(r^2e^{(\lambda+\nu)/2}F^{01}\right)}{dr}=4\pi
r^2e^{\lambda/2}\rho,
\end{equation}
whose integration yields
\begin{eqnarray}\nonumber
F^{01}=\frac{e^{-(\lambda+\nu)/2}Q(t,r)}{r^2},
\end{eqnarray}
where $Q(t,r)=4\pi\int_{0}^r r^2\rho e^{\lambda/2}dr$ is the total
amount of charge within the sphere.

The energy-momentum tensor follows the conservation identity
$T_{~;j}^{ij}=0$, which governs hydrodynamics of the fluid and leads
to the following relations
\begin{eqnarray}\label{15}
\frac{\partial T_{0}^0}{\partial t}+\frac{\partial T_{1}^0}{\partial
r}+\frac{1}{2}\left(\frac{4}{r}+\frac{\partial}{\partial
r}(\lambda+\nu)\right)T_{0}^1+\frac{1}{2}\left(T_{0}^0-
T_{1}^1\right)\frac{\partial\lambda}{\partial t}=0,
\\\label{16}\frac{\partial T_{1}^0}{\partial
t}+\frac{\partial T_{1}^1}{\partial r}+\frac{1}{2}\left(T_{1}^1-
T_{0}^0\right)\frac{\partial\nu}{\partial
r}-\frac{2}{r}\left(p-T_{1}^1\right)+\frac{1}{2}T_{1}^0
\frac{\partial}{\partial t}(\lambda+\nu)=0,
\end{eqnarray}
where $T_{0}^1=-e^{\nu-\lambda}T_{1}^0$. The non-zero components of
energy-momentum tensor are
\begin{eqnarray}\nonumber
T^0_{0}=-\sigma-\frac{Q^2}{8\pi r^4}, \quad
T^1_{1}=p-\frac{Q^2}{8\pi r^4}, \quad
T^2_{2}=T^3_{3}=p+\frac{Q^2}{8\pi r^4}.
\end{eqnarray}
All the quantities governing the motion remain independent of time
during the state of hydrostatic equilibrium. The surface stresses
describing equilibrium state are denoted by zero subscript. In this
context, Eqs.(\ref{6}), (\ref{7}) and (\ref{16}) take the form
\begin{eqnarray}\label{19}
\frac{d}{dr}(re^{-\lambda_{0}})&=&1-\frac{8\pi G
r^2}{c^4}\sigma_{0}-\frac{GQ^2}{c^4r^2}, \\\label{20}
\frac{1}{r}e^{-\lambda_{0}}\frac{d\nu_{0}}{dr}&=&\frac{1}{r^2}
(1-e^{-\lambda_{0}})+\frac{8\pi Gp_{0}}{c^4}-\frac{GQ^2}{c^4r^4},
\\\label{21} \frac{dp_{0}}{dr}&=&\frac{1}{2}\frac{d\nu_{0}}{dr}(p_{0}
+\sigma_{0})+\frac{1}{8\pi}\frac{d}{dr}
\left(\frac{Q^2}{r^4}\right)+\frac{Q^2}{4\pi r^5}.
\end{eqnarray}
Following Eqs.(\ref{6}) and (\ref{7}), we also have a useful
relation
\begin{equation}\label{22}
\frac{e^{-\lambda_{0}}}{r}\frac{d}{dr}(\lambda_{0}+
\nu_{0})=(p_{0}+\sigma_{0})\frac{8\pi G}{c^4}.
\end{equation}
We take the Reissner-Nordstr\"{o}m (RN) spacetime in the exterior
region as
\begin{eqnarray}\nonumber
ds^2&=&-\left(1-\frac{2GM}{rc^2}+\frac{GQ^2}{r^2c^4}\right)
dt^2+\left(1-\frac{2GM}{rc^2}+\frac{GQ^2}{r^2c^4}\right)^{-1}
dr^2\\\label{22a}&+&r^2(d\theta^2+\sin^2\theta d\phi^2),
\end{eqnarray}
where $M$ corresponds to the total mass of the sphere. The
hydrostatic equilibrium describes the state of fluid in which
pressure gradient force is balanced by the gravitational force. When
one of these forces overcome the other, the stability of the system
is disturbed leading to an unstable system. The equation describing
hydrostatic equilibrium is obtained by eliminating $\nu_{0}$ from
Eqs.(\ref{20}) and (\ref{21}) as
\begin{eqnarray}\nonumber
&&\left(1-\frac{2GM_{r}}{rc^2}+\frac{GQ^2}{r^2c^4}\right)
\frac{dp_{0}}{dr}=-\frac{1}{c^2}\left(\frac{GM_{r}}{r^2}-
\frac{GQ^2}{r^3c^2}+\frac{4\pi
G}{c^2}pr\right)\\\label{23}&&\times(p_{0}+\sigma_{0})+
\frac{1}{4\pi}\left(1-\frac{2GM_{r}}{rc^2}+\frac{GQ^2}
{r^2c^4}\right)\left(\frac{Q^2}{r^5}+\frac{1}{2}\frac{d}
{dr}\left(\frac{Q^2}{r^5}\right)\right),
\end{eqnarray}
where the left and right hand sides correspond to pressure gradient
and gravitational terms, respectively and
\begin{equation}\label{24}
M_{r}=\frac{4\pi G}{c^4}\int_{0}^r \sigma_{0}r^2
dr+\frac{G}{2c^4}\int_{0}^r \frac{Q^2}{r^2}dr,
\end{equation}
is the Misner-Sharp mass function.

\section{Equations Governing Radial Oscillations}

Here we discuss the motion of gaseous masses undergoing radial
oscillations. The non-zero components of four-velocity are given by
\begin{equation}\label{25}
u^0=e^{-\frac{\nu_{0}}{2}}, \quad u_{0}=-e^{\frac{\nu_{0}}{2}},\quad
u^1=ve^{-\frac{\nu_{0}}{2}}, \quad
u_{1}=ve^\frac{\lambda_{0}-\nu_{0}}{2},
\end{equation}
where $v=\frac{dr}{dt}$ is the radial velocity component. These
components can be calculated with respect to spacetime coordinates
by $u^i=\frac{dx^i}{ds}$. The stability of any gaseous mass under
perturbation ultimately gives rise to the dynamical evolution of
gravitating system. We assume that an equilibrium configuration is
perturbed such that it does not affect the spherical symmetry. We
consider only linear terms so that the respective values in the
perturbed state become
\begin{equation}
\lambda=\lambda_{0}+\delta\lambda,\quad \nu=\nu_{0}+\delta\nu, \quad
p=p_{0}+\delta p, \quad \sigma=\sigma_{0}+\delta\sigma, \quad
Q=Q_{0}+\delta Q.
\end{equation}
We follow the Eulerian approach \cite{c} for perturbations such that
the corresponding linearized forms (governing the radial
perturbations) through Eqs.(\ref{19}) and (\ref{20}) are
\begin{eqnarray}\label{30}
\frac{\partial}{\partial
r}(re^{-\lambda_{0}}\delta\lambda)=\frac{2G}{c^4}\left(4\pi
r^2\delta\sigma-\frac{Q_{0}\delta Q}{r^2}\right), \\\label{31}
\frac{e^{-\lambda_{0}}}{r}\left[\frac{\partial}{\partial
r}\delta\nu-\delta\lambda\frac{d\nu_{0}}{dr}\right]=
\frac{1}{r^2}e^{-\lambda_{0}}\delta\lambda+\frac{8\pi G}{c^4}\delta
p-\frac{2GQ_{0}\delta Q}{c^4r^4},
\end{eqnarray}
here $\delta\lambda$, $\delta\nu$, $\delta\sigma$, $\delta p$ and
$\delta Q$ represent the Eulerian changes. Equations (\ref{8}) and
(\ref{16}) can be written appropriately in linearized forms as
\begin{eqnarray}\label{32}
\frac{e^{-\lambda_{0}}}{r}\frac{\partial}{\partial
t}\delta\lambda=-\frac{4\pi
G}{c^4}\left(2(p_{0}+\sigma_{0})v-\frac{Q_{0}\delta Q}{r^4}
\right),\\\nonumber
(p_{0}+\sigma_{0})e^{\lambda_{0}-\nu_{0}}\frac{\partial v}{\partial
t}+\frac{\partial}{\partial r}\delta
p+\frac{1}{2}(p_{0}+\sigma_{0})\frac{\partial}{\partial
r}\delta\nu\\\label{33}
+\frac{1}{2}(\delta
p+\delta\sigma)\frac{d\nu_{0}}{dr}+\frac{1}{8\pi}\frac{Q_{0} \delta
Q}{r^4}-\frac{1}{4\pi}\frac{\partial}{\partial
r}\left[\frac{Q_{0}\delta Q}{r^4}\right]=0.
\end{eqnarray}

Let us introduce a Lagrangian displacement $``\eta"$ such that
$v=\frac{\partial\eta}{\partial t}$. Integration of Eq.(\ref{32})
leads to
\begin{equation}\label{35}
\frac{e^{-\lambda_{0}}}{r}\delta\lambda=-\frac{8\pi
G}{c^4}(p_{0}+\sigma_{0})\eta+\frac{4\pi GQ_{0}}{c^4r^4}
\int\delta
Qdt.
\end{equation}
Using Eq.(\ref{22}), this equation takes the form
\begin{equation}\label{36}
\delta\lambda=-\frac{d}{dr}(\lambda_{0}+\nu_{0})\eta+\frac{4\pi
GQ_{0}e^{\lambda_{0}}}{c^4r^3}\int\delta Q dt.
\end{equation}
Solving Eq.(\ref{30}) and (\ref{35}), it follows that
\begin{equation}\label{37}
\delta\sigma=\frac{1}{r^2}\frac{\partial}{\partial
r}\left[-r^2(\sigma_{0}+p_{0})\eta+\frac{Q^2}{2r^4}\int\delta
Qdt\right]+\frac{Q_{0}}{4\pi r^4}\delta Q,
\end{equation}
which yields
\begin{eqnarray}\nonumber
\delta\sigma&=&-\eta\frac{dp_{0}}{dr}-\eta\frac{d\sigma_{0}}
{dr}-\frac{1}{r^2}(p_{0}+\sigma_{0})\frac{\partial} {\partial
r}(\eta r^2)+\frac{1 }{r^2}\frac{\partial}{\partial
r}\left[\frac{Q_{0}}{2r^4}\int\delta Q
dt\right]\\\label{38}
&+&\frac{Q_{0}}{4\pi r^4}\delta Q.
\end{eqnarray}
Using Eq.(\ref{21}), it follows that
\begin{eqnarray}\nonumber
\delta\sigma&=&-\eta\frac{d\sigma_{0}}{dr}-\frac{e^{\nu_{0}/2}}
{r^2}(p_{0}+\sigma_{0})\frac{\partial}{\partial r}\left[\eta
r^2e^{-{\nu_{0}/2}}\right]-\frac{\eta}{8\pi}\frac{d}{d
r}\left[\frac{Q^2}{r^4}\right]\\\label{39} &+&\frac{1}{r^2}
\frac{\partial}{\partial r}\left[\frac{Q_{0}}{2r^4}\int\delta Q
dt\right]+\frac{Q_{0}}{4\pi r^4}\delta Q.
\end{eqnarray}
Substituting $\delta\lambda$ from Eq.(\ref{35}) in (\ref{31}), we
obtain
\begin{eqnarray}\nonumber
\frac{e^{-\lambda_{0}}}{r}\frac{\partial}{\partial
r}\delta\nu&=&\left[(p_{0}+\sigma_{0})\left(\frac{1}{r}+
\frac{d\nu_{0}}{dr}\right)\eta+\delta p\right]\frac{8\pi
G}{c^4}\\\label{40}&+&\frac{4\pi
GQ_{0}}{c^4r^4}\left[\frac{d\nu_{0}}{dr}-\frac{\eta}{r}
\right]\int\delta Q dt-\frac{2GQ_{0}}{c^4r^4}\delta Q,
\end{eqnarray}
which in accordance of Eq.(\ref{22}) leads to
\begin{eqnarray}\nonumber
(p_{0}+\sigma_{0})\frac{\partial}{\partial
r}\delta\nu&=&\frac{d}{dr}(\lambda_{0}+\nu_{0})\left\{\left [\delta
p-(p_{0}+\sigma_{0})\left(\frac{1}{r}+\frac{d\nu_{0}}{dr}
\right)\eta
\right]\right.\\\label{41}&+&\left.\frac{Q_{0}}{2r^4}\left
[\frac{d\nu_{0}}{dr}-\frac{\eta}{r}\right] \int\delta Q
dt-\frac{Q_{0}}{4\pi r^4}\delta Q\right\}.
\end{eqnarray}
Now we assume time dependent perturbations in the form of Lagrangian
displacement, i.e., $\eta e^{i\omega t}$, where $\omega$ is the
characteristic frequency to be evaluated. The Lagrangian
displacement $\eta$ connects the fluids elements in equilibrium with
corresponding one in the perturbed configuration. Since the
equations have natural modes of oscillations, so they will depend on
time. Considering $\delta\lambda,~\delta\nu,$ $\delta
p,~\delta\sigma$ and $\delta Q$ as time dependent amplitudes of the
respective quantities, Eq.(\ref{33}) with (\ref{41}) can be
rewritten as
\begin{eqnarray}\nonumber
\omega^2e^{\lambda_{0}-\nu_{0}}(p_{0}+\sigma_{0})\eta&=&\delta p
\frac{d}{dr}\left(\nu_{0}+\frac{1}{2}\lambda_{0}\right)+\frac{d}
{dr}\delta
p+\frac{1}{2}\delta\sigma\frac{d\nu_{0}}{dr}-\frac{1}{2}(p_{0}+
\sigma_{0})\\\nonumber &\times&\left(\frac{d\nu_{0}}{dr}+\frac
{d\lambda_{0}}{dr}\right)\left(\frac{1}{r}+\frac{d\nu_{0}}{dr}
\right)\eta+\frac{Q_{0}}{8\pi
r^4}\frac{d}{dr}(\lambda_{0}+\nu_{0})\\\nonumber &\times&\left
\{2\pi\left(\frac{d\nu_{0}}{dr}- \frac{\eta}{r}\right)\int\delta Q
dt-\delta Q\right\}+\frac{1}{4\pi}\left\{\frac{Q_{0}\delta
Q}{2r^4}\right.\\\label{43} &-&\left.\frac{d}{dr}\left(\frac{Q_{0}
\delta Q}{r^4}\right)\right\}.
\end{eqnarray}

\section{The Conservation of Baryon Number}

In order to discuss the perturbed state of pressure in terms of
Lagrangian displacement $\eta$, an additional assumption is required
which can relate physical aspects of relativistic theory with the
gaseous mass undergoing adiabatic radial oscillations. In this
context, the required supplementary condition can be satisfied by
conservation of baryon number in the framework of GR as
$(Nu^j)_{;j}=0$, or
\begin{equation}\label{45}
\frac{\partial}{\partial x^j}(Nu^j)+Nu^j\frac{\partial}{\partial
x^j}\ln\sqrt{-g}=0,
\end{equation}
where $N$ is the baryon number per unit volume. The conservation of
baryon number plays a vital role in collecting different models of
the universe. According to this law, the number of particles may
vary but their total number will remain conserved during the fluid
flow. This change occurs due to loss or gain of net fluxes. Here we
consider fluid obeying this identity. Equation (\ref{45}) through
(\ref{25}) leads to
\begin{eqnarray}\nonumber
&&\frac{\partial}{\partial
t}(Ne^{-\nu_{0}/2})+\frac{\partial}{\partial
r}(Nve^{-\nu_{0}/2})+Nve^{-\nu_{0}/2}\frac{\partial}{\partial
r}\left(\frac{2}{r}+\frac{1}{2}[\nu+\lambda]\right)\\\label{46}&&
+\frac{N}{2}e^{-\nu_{0}/2}\frac{\partial}{\partial
t}(\nu+\lambda)=0.
\end{eqnarray}

We assume the perturbation
\begin{equation}\label{47}
N=N_{0}(r)+\delta N(r,t),
\end{equation}
keeping only the linear terms in $v$, Eq.(\ref{46}) takes the form
\begin{eqnarray}\nonumber
&&\frac{1}{r^2}\frac{\partial}{\partial
r}(N_{0}r^2ve^{-\nu_{0}/2})+e^{-\nu_{0}/2}\frac{\partial}{\partial
t}\delta
N+\frac{1}{2}N_{0}e^{-\nu_{0}/2}\frac{d}{dr}(\nu_{0}+\lambda_{0})
\\\label{48}&&+\frac{1}{2}N_{0}e^{-\nu_{0}/2}\frac{\partial}
{\partial t}\delta\lambda=0,
\end{eqnarray}
whose integration in terms of Lagrangian displacement $\eta$ leads
to
\begin{equation}\label{49}
\delta N+\frac{N_{0}}{2}\left[\eta\frac{d}{dr}(\nu_{0}+\lambda_{0})
+\delta\lambda\right]+\frac{1}{r^2}e^{\nu_{0}/2}\frac{\partial}
{\partial r}\left(N_{0}r^2\eta e^{-\nu_{0}/2}\right)=0.
\end{equation}
Using Eq.(\ref{36}), it follows that
\begin{equation}\label{51}
\delta N=-\eta\frac{dN_{0}}{dr}-\frac{N_{0}}{r^2}e^{\nu_{0}/2}
\frac{\partial}{\partial r}\left(r^2\eta
e^{-\nu_{0}/2}\right)+\frac{2\pi
GN_{0}Q_{0}}{c^4r^3}e^{\lambda_{0}}\int\delta Qdt.
\end{equation}
We consider an equation of state in the form
\begin{equation}
N=N(\sigma,p),
\end{equation}
so that Eqs.(\ref{39}) and (\ref{51}) together give
\begin{equation}\label{53}
\delta p=-\eta\frac{dp_{0}}{dr}-p_{0}\Gamma\frac{e^{\nu_{0}/2}}
{r^2}\frac{\partial}{\partial r}\left(r^2\eta
e^{-\nu_{0}/2}\right)+\frac{\alpha Q_{0}}{r^3}\int\delta Qdt,
\end{equation}
where
\begin{equation}
\alpha=\frac{1}{\partial N/\partial p}\left\{\frac{2\pi
GN_{0}}{c^4}e^{\lambda_{0}}-\frac{1}{2r}\frac{dN} {d\sigma}\right\},
\end{equation}
and $\Gamma$ is the adiabatic index (ratio of specific heats)
defined by
\begin{equation}\label{54}
\Gamma=\frac{1}{p\partial N/\partial
p}\left\{N-(\sigma+p)\frac{\partial N}{\partial\sigma}\right\}.
\end{equation}
This relates the pressure and density fluctuations and measures the
stiffness of the fluid.

\section{Pulsation Equation and Variational Principle}

The linear pulsation corresponds to the oscillation frequencies and
different modes of small perturbations applied to equilibrium
spherical configuration. Inserting the values of $\delta\sigma$ and
$\delta p$ from Eqs.(\ref{37}) and (\ref{53}) in (\ref{43}), it
follows that
\begin{eqnarray}\nonumber
&&\omega^2e^{\lambda_{0}-\nu_{0}}(p_{0}+\sigma_{0})\eta
=-\eta\left(\frac{d\nu_{0}}{dr}+\frac{1}{2}\frac
{d\lambda_{0}}{dr}\right)\frac{dp_{0}}{dr}-\frac{d}{dr}
\left(\eta\frac{dp_{0}}{dr}\right)-\frac{1}{2}\left
\{\frac{2}{r}\right.\\\nonumber&&\times\left.(p_{0}+
\sigma_{0})\eta+\frac{d}{dr}[(p_{0}+\sigma_{0})\eta]
\right\}-\frac{1}{2}\eta(p_{0}+\sigma_{0})\left(\frac
{d\nu_{0}}{dr}+\frac{d\lambda_{0}}{dr}\right)\left
(\frac{1}{r}+\frac{d\nu_{0}}{dr}\right)\\\nonumber&&-
e^{-(\nu_{0}+\frac{\lambda_{0}}{2})}\frac{d}{dr}\left
\{e^{(\nu_{0}+\frac{\lambda_{0}}{2})}\Gamma
p_{0}\frac{e^{\nu_{0}/2}}{r^2}\frac{d}{dr}(r^2\eta
e^{-\nu_{0}/2})\right\}+e^{-\lambda_{0}/2}\frac{d}{dr}
\left\{\frac{\beta Q_{0}}{r^3}\right.\\\nonumber&&\times
\left.e^{\lambda_{0}/2}\int\delta
Qdt\right\}+\frac{Q_{0}}{r^3}\frac{d\nu_{0}}{dr}\int\delta
Qdt\left\{\beta+\frac{1}{4r}+\frac{1}{4\pi
r}\left(\frac{d\nu_{0}}{dr}-\frac{\eta}{r}\right)\right\}
\\\nonumber&&+\frac{Q_{0}}{4\pi r^4}\frac{d\lambda_{0}}{dr}
\left\{\frac{d\nu_{0}}{dr}-\frac{\eta}{r}\int\delta Qdt-
\frac{\delta Q}{2}\right\}+\frac{Q_{0}}{8\pi r^4}\delta
Q+\frac{e^{\nu_{0}/2}}{4\pi}\frac{d}{dr}\left\{\frac{Q_{0}
e^{\nu_{0}/2}\delta Q}{r^4}\right\}.\\\label{55}
\end{eqnarray}
Substituting $\frac{dp_{0}}{dr}$ from Eq.(\ref{21}) in the above
equation, we have
\begin{eqnarray}\nonumber
&&\omega^2e^{\lambda_{0}-\nu_{0}}(p_{0}+\sigma_{0})\eta
=\frac{1}{2}(p_{0}+\sigma_{0})\eta\left\{\frac
{d^2\nu_{0}}{dr^2}-\frac{3}{r}\frac{d\nu_{0}}{dr}-
\frac{1}{r}\frac{d\lambda_{0}}{dr}-\frac{1}{2}\frac
{d\lambda_{0}}{dr}\frac{d\nu_{0}}{dr}\right\}
\\\nonumber&&-\frac{5}{2\pi}\frac{Q^2}{r^6}-\frac{1}{8\pi
r^4}\frac{d^2}{dr^2}(Q^2)+e^{-\lambda_{0}}\frac{d}{dr}
\left\{\frac{\beta
Q_{0}}{r^3}e^{-\lambda_{0}/2}\widetilde{Q}\right\}+
\frac{Q_{0}\widetilde{Q}}{r^3}\frac{d\nu_{0}}{dr}
\left\{\beta+\frac{1}{4\pi}\right.\\\nonumber&&\left.+ \frac{1}{4\pi
r}\left(\frac{d\nu_{0}}{dr}-\frac{\eta}{r}\right)\right\}
+\frac{Q_{0}}{4\pi
r^4}\frac{d\lambda_{0}}{dr}\left\{\left(\frac{d\nu_{0}}{dr}
-\frac{\eta}{r}\right)\widetilde{Q}-\frac{\delta
Q}{2}\right\}+\frac{Q_{0}\delta Q}{8\pi
r^4}\\\label{56}&&+\frac{e^{\nu_{0}/2}}{4\pi}\frac{d}{dr}
\left\{\frac{Q_{0}e^{-\nu_{0}/2}\delta Q}{r^4}\right\},
\end{eqnarray}
where $\int\delta Qdt=\widetilde{Q}$. Under the equilibrium
condition, Eq.(\ref{8}) yields
\begin{equation}\label{57}
\left\{\frac{16\pi
Gp_{0}}{c^4}+\frac{2GQ^2_{0}}{c^4r^4}\right\}e^{\lambda_{0}}
=\frac{d^2\nu_{0}}{dr^2}+\frac{1}{r}\frac{d}{dr}(\nu_{0}-
\lambda_{0})+\frac{1}{2}\left(\frac{d\nu_{0}}{dr}\right)^2-
\frac{1}{2}\frac{d\lambda_{0}}{dr}\frac{d\nu_{0}}{dr}.
\end{equation}
Using this expression and Eq.(\ref{21}), Eq.(\ref{56}) takes the
form
\begin{eqnarray}\nonumber
&&\omega^2e^{\lambda_{0}-\nu_{0}}(p_{0}+\sigma_{0})\eta
=\frac{4}{r}\frac{dp_{0}}{dr}\eta-\frac{1}{p_{0}+\sigma_{0}}
\left(\frac{dp_{0}}{dr}\right)^2\eta+\frac{8\pi
Gp_{0}}{c^4}e^{\lambda_{0}}\\\nonumber&&\times
(p_{0}+\sigma_{0})\eta+\frac{d}{dr}\left[e^{\lambda{0}
+3\nu_{0}/2}\frac{p_{0}\Gamma}{r^2}\frac{d}{dr}\left(r^2\eta
e^{-\nu_{0}/2}\right)\right]e^{-(\nu_{0}+\lambda_{0}/2)}
\\\nonumber&&+\frac{1}{p_{0}+\sigma_{0}}\frac{dp_{0}}{dr}
\eta\left(\frac{1}{4\pi r^4}\frac{d(Q^2)}{dr}-\frac{Q^2}{\pi
r^5}\right)+\frac{dQ^2}{dr}\frac{\eta}{2\pi
r^5}\left[\frac{1}{p_{0}+\sigma_{0}}\frac{Q^2}{4\pi
r^4}-1\right]\\\nonumber&&-\frac{\eta}{p_{0}+\sigma_{0}}
\frac{1}{(2\pi
r^4)^2}\left[\frac{Q^4}{r^2}+\frac{1}{16}\left(\frac{dQ^2}
{dr}\right)^2\right]-\frac{5Q^2}{2\pi r^6}-\frac{1}{8\pi
r^4}\frac{d^2}{dr^2}(Q^2)\\\nonumber&&+\frac{GQ_{0}^2}{c^4r^4}
(p_{0}+\sigma_{0})\eta
e^{\lambda_{0}}+e^{-\lambda_{0}}\frac{d}{dr}\left (\frac{\beta
Q_{0}\widetilde{Q}}{r^3}e^{\lambda_{0}/2}\right)+\frac{Q_{0}
\widetilde{Q}}{r^3}\frac{d\nu_{0}}{dr}\left[\beta+\frac{1}{4\pi}
\right.\\\nonumber&&\left.+\frac{1}{4\pi
r}\left(\frac{d\nu_{0}}{dr}-\frac{\eta}{r}\right)\right]
+\frac{Q_{0}\delta Q}{8\pi
r^4}\left(1-\frac{d\lambda_{0}}{dr}\right)+\frac{e^{\nu_{0}/2}}
{4\pi}\frac{d}{dr}\left(\frac{Q_{0}\delta
Q}{r^4}e^{-\nu_{0}/2}\right).\\\label{59}
\end{eqnarray}
This is the required pulsation equation which satisfies the boundary
conditions, i.e., $\eta=0$ at $r=0$ and $\delta p=0$ at $r=R$. This
constitutes a characteristic value problem for $\omega^2$ obtained
by multiplying the equation with $\eta
r^2e^{(\lambda_{0}+\nu_{0})/2}$ and integrating over values of $r$
as
\begin{eqnarray}\nonumber
&&\omega^2\int_{0}^R
e^{(3\lambda-\nu)/2}r^2\eta^2(p+\sigma)dr=\int_{0}^R
e^{(\lambda+3\nu)/2}\frac{p\Gamma}{r^2}\left[\frac{d}{dr}
\left(r^2\eta e^{-\nu/2}\right)\right]^2dr\\\nonumber&&+ \frac{8\pi
G}{c^4}\int_{0}^Re^{(3\lambda+\nu)/2}pr^2\eta^2(p+\sigma)dr
-\int_{0}^R\frac{r^2\eta^2}{p+\sigma}e^{(\lambda+\nu)/2}
\left(\frac{dp}{dr}\right)^2dr\\\nonumber&&+4\int_{0}^R
r\eta^2e^{(\lambda+\nu)/2}\frac{dp}{dr}dr+\int_{0}^R
\frac{\eta^2}{p+\sigma}e^{(\lambda+\nu)/2}\frac{dp}{dr}
\left(\frac{1}{4\pi}\frac{dQ^2}{dr}-\frac{Q^2}{\pi
r^3}\right)dr\\\nonumber&&+\int_{0}^R \frac{\eta^2}{2\pi
r^3}e^{(\lambda+\nu)/2}\frac{dQ^2}{dr}\left(\frac{1}{p+\sigma}
\frac{Q^2}{4\pi r^4}-1\right)dr-\frac{5}{2\pi}\int_{0}^R \frac{\eta
Q^2}{r^4}e^{(\lambda+\nu)/2}dr \\\nonumber&&-\int_{0}^R
\frac{\eta^2}{p+\sigma}\frac{e^{(\lambda+\nu)/2}}{\left(2\pi
r^3\right)^2}\left[\left(\frac{Q^2}{r}\right)^2+\frac{1}{16}\left
(\frac{dQ^2}{dr}\right)^2\right]dr+\int_{0}^R
\frac{GQ_{0}^2\eta^2}{c^4r^4}
\\\nonumber&&\times(p+\sigma) e^{(3\lambda+\nu)/2}dr-\int_{0}^R
\frac{\eta e^{(\lambda+\nu)}} {8\pi
r^2}\frac{d^2}{dr^2}\left(Q^2\right)dr+\int_{0}^R r^2\eta
e^{(\nu-\lambda)/2}
\end{eqnarray}
\begin{eqnarray}\nonumber
&&\times\frac{d}{dr}\left(\frac{\beta
Q_{0}\widetilde{Q}}{r^3}e^{\lambda_{0}/2}\right)dr+\int_{0}^R
\frac{\eta Q_{0}\delta Q}{8\pi r^2}
e^{(\nu+\lambda)/2}\left(1-\frac{d\lambda_{0}}{dr}\right)dr
\\\nonumber&&+\int_{0}^R \frac{\eta Q_{0}\widetilde{Q}}{r}
e^{(\nu+\lambda)/2}\frac{d\nu_{0}}{dr}\left[\beta+\frac{1}{4\pi}
+\frac{1}{4\pi
r}\left(\frac{d\nu_{0}}{dr}-\frac{\eta}{r}\right)\right]dr
\\\label{61}&&+\int_{0}^R\frac{\eta r^2}{4\pi}
e^{(\nu+\lambda)/2}\frac{d}{dr}\left(\frac{Q_{0}\delta
Q}{r^4}e^{-\nu_{0}/2}\right)dr.
\end{eqnarray}
The corresponding orthogonality condition is defined as
\begin{equation}
\int_{0}^R e^{(3\lambda-\nu)/2}r^2(p+\sigma)\eta^{(i)}\eta^{(j)}=0,
\quad (i\neq j),
\end{equation}
where $\eta^{(i)}$ and $\eta^{(j)}$ give proper solutions associated
with different characteristic values of $\omega^2$. To investigate
dynamical instability of spherical star, the right-hand side of
Eq.(\ref{61}) should vanish by choosing a trial function $\xi$
satisfying the given boundary conditions. In the following, we
discuss the conditions for dynamical instability by taking two
special models.

\subsection{The Homogeneous Model of Sphere}

First we consider the homogeneous sphere with constant energy
density $\sigma$ and study the conditions for its dynamical
instability. Equations (\ref{23}) and (\ref{24}) governing the
hydrostatic equilibrium allow the integration \cite{b} such that we
can write
\begin{equation}
x^2=1-\frac{r^2}{a^2}+\frac{b^2}{r^2}, \quad
x_{1}^2=1-\frac{R^2}{a^2}+\frac{b^2}{R^2},
\end{equation}
where $a^2=\frac{3c^4}{8\pi G\sigma}$ and
$b^2=\frac{2GQ^2}{c^4}$.
The solutions of the relevant physical quantities
can be determined in terms of $x$ and $x_{1}$ as
\begin{eqnarray}\label{65a}
p=\sigma\frac{x-x_{1}}{3x_{1}-x}, \quad
e^{\nu}=\frac{1}{4}[3x_{1}-x]^2, \quad
e^{\lambda}=\frac{1}{x^2}.
\end{eqnarray}
The necessary condition for the positivity of pressure yields
$3x_{1}>1$ which leads to
\begin{equation}\nonumber
\frac{R^2}{a^2}-\frac{b^2}{R^2}<\frac{8}{9}.
\end{equation}
Using the inertial mass, this takes the form
\begin{equation}\label{69}
R>\frac{9}{8}\left(\frac{2GM}{c^2}-\frac{GQ^2}{Rc^4}\right)
=\frac{9}{8}R_{N},
\end{equation}
where $R_{N}$ is RN radius. Inserting the physical quantities in
Eq.(\ref{61}), it follows that
\begin{eqnarray}\nonumber
&&4a^2\omega^2x_{1}\int_{0}^{\xi_{1}}\frac{\xi^2\eta^2}
{x^3(3x_{1}-x)^2}d\xi=x_{1}\int_{0}^{\xi_{1}}
\frac{2x^2-9x_{1}^2-1}{x^3(3x_{1}-x)^2}\xi^2\eta^2d\xi
\\\nonumber&+&\frac{\Gamma}{8}\int_{0}^{\xi_{1}}(x-x_{1})
(3x_{1}-x)^2\frac{1}{x\xi}\left[\frac{d}{d\xi}(\eta\xi^2
e^{-\nu/2})\right]^2d\xi+\frac{1}{16\pi a^3x_{1}}
\\\nonumber&\times&\int_{0}^{\xi_{1}}\frac{\eta^2}{x\xi^2}
\frac{dp}{d\xi}\left(\frac{dQ^2}{d\xi}-\frac{4Q^2}{\xi}\right)
d\xi-\frac{5}{4\pi
a^3}\int_{0}^{\xi_{1}}\frac{Q^2\eta}{\xi^4}\frac{3x_{1}-x}
{x}d\xi\\\nonumber&+&\frac{1}{4\pi
a^3}\int_{0}^{\xi_{1}}\frac{\eta^2(3x_{1}-x)}{x\xi^2}
\left[\frac{Q^2(3x_{1}-x)}{8\pi
a^4x_{1}\xi^4}-1\right]d\xi-\frac{1}{16\pi
ax_{1}}\\\nonumber&\times&\int_{0}^{\xi_{1}}
\frac{\eta^2(3x_{1}-x)^2}{x(2\pi
a^3\xi^3)^2}\left(\frac{dQ^2}{d\xi}-\frac{4Q^2}{\xi}\right)
d\xi+\frac{Gx_{1}}{c^4}\int_{0}^{\xi_{1}}\frac{\eta^2Q_{0}^2}
{x^3\xi^2}d\xi\\\nonumber&-&\frac{1}{32\pi
a^3}\int_{0}^{\xi_{1}}\frac{\eta(3x_{1}-x)^2}{x^2\xi^2}
\frac{d^2}{d\eta^2}(Q^2)d\xi+\frac{1}{2a}\int_{0}^{\xi_{1}}
x(3x_{1}-x)\eta\xi^2\\\nonumber&\times&\frac{d}{d\xi}
\left[\frac{\beta
Q_{0}\widetilde{Q}}{\xi^3}e^{-\lambda/2}\right]d\xi+
\frac{1}{2a^2}\int_{0}^{\xi_{1}}\frac{\eta
Q_{0}\widetilde{Q}}{\xi}\frac{3x_{1}-x}{x}\frac{d\nu_{0}}
{d\xi}\left[\beta+\frac{1}{4\pi\xi}\right.\\\nonumber
&\times&\left.\left(\frac{d\nu_{0}}{d\xi}-
\frac{\eta}{\xi}\right)\right]d\xi+\frac{1}{16\pi
a^2}\int_{0}^{\xi_{1}}\frac{\eta Q_{0}\delta
Q}{\xi^2}\frac{3x_{1}-x}{x}\left(1-\frac{1}{a}\frac
{d\lambda_{0}}{d\xi}\right)d\xi\\\label{70}&+&\frac{1}{4\pi
a^2}\int_{0}^{\xi_{1}}\frac{\eta\xi^2(3x_{1}-x)}{x}
\frac{d}{d\xi}\left(\frac{Q_{0}\delta Q}{\xi^4(3x_{1}-x)}
\right)d\xi,
\end{eqnarray}
where $\xi=\frac{r}{a}$, $\xi_{1}=\frac{R}{a}-\frac{b}{R}$ and
$\Gamma$ is assumed to be constant.

We take the trial function as
\begin{equation}
\eta=\xi e^{\nu/2}=\frac{1}{2}\xi(x_{1}-x),
\end{equation}
for which Eq.(\ref{70}) becomes
\begin{eqnarray}\nonumber
&&(a\omega)^{2}x_{1}\int_{0}^{\xi_{1}}\frac{\xi^{4}}
{x^{3}}d\xi=\frac{1}{4}x_{1}\int_{0}^{\xi_{1}}
(2x^{2}-1-9x_{1}^{2})\frac{\xi^{4}}{x^{3}}d\xi+
\frac{9}{8}\int_{0}^{\xi_{1}}(x-x_{1})\\\nonumber&\times&
(3x_{1}-x)^{2}\frac{\xi^{2}}{x}d\xi+\frac{1}{4a^{3}xx_{1}}\int_{0}
^{\xi_{1}}(3x_{1}-x)^{4}\left(-\frac{Q^{2}}{\pi\xi}\right)
\frac{d}{d\xi}\left(\frac{x-x_{1}}{3x_{1}-x}
\right)d\xi\\\nonumber&-&\frac{1}{(4\pi a^{3})^{2}x_{1}}
\int_{0}^{\xi_{1}}\frac{(3x_{1}-x)^{2}}{x}
\left(\frac{Q^{2}}{\xi}\right)^{2}d\xi+\frac{Gx_{1}}{4c^{4}}
\int_{0}^{\xi_{1}}Q_{0}^{2}\frac{(3x_{1}-x)^{2}}{\xi
x^{3}}d\xi
\end{eqnarray}
\begin{eqnarray}\nonumber
&-& \frac{5}{8\pi
a^{3}}\int_{0}^{\xi_{1}}\frac{Q^{2}(3x_{1}-x)^{2}}{x\xi^{3}}
d\xi+\frac{1}{4a}\int_{0}^{\xi_{1}}x\xi^{3}(3x_{1}-x)
\frac{d}{d\xi}\left(\frac{3Q_{0}
\widetilde{Q}}{x\xi^{3}}\right)d\xi\\\nonumber&+&\frac{1}{32\pi
a^{2}}\int_{0}^{\xi^{1}}\frac{Q_{0}\delta
Q}{x\xi}(3x_{1}-x)^{2}d\xi+\frac{1}{8\pi
a^{2}}\int_{0}^{\xi_{1}}\frac{\xi^{3}}{x}(3x_{1}-x)^{2}
\\\label{73}&\times&\frac{d}{d\xi}\left(\frac{Q_{0}\delta
Q}{(3x_{1}-x)\xi^{4}}\right)d\xi.
\end{eqnarray}
Substituting $x=\cos\theta$ and $\xi=\sin\theta$ in the above
equation, we obtain
\begin{eqnarray}\nonumber
&&(a\omega)^{2}\cos\theta_{1}\int_{0}^{\theta_{1}}
\frac{\sin^{4}\theta}{\cos^{2}\theta}d\theta=
\frac{\cos\theta_{1}}{4}\int_{0}^{\theta_{1}}
(2\cos^{2}\theta-1-9\cos^{2}\theta_{1})\frac{\sin^{4}
\theta}{\cos^{2}\theta}d\theta\\\nonumber&+&\frac{9}{8}\Gamma
\int_{0}^{\theta_{1}}(\cos\theta-\cos\theta_{1})
(3\cos\theta_{1}-\cos\theta)^{2}\sin^{2}\theta
d\theta-\frac{1}{4a^{3}\cos\theta_{1}}\\\nonumber&\times&\int_{0}
^{\theta_{1}}\frac{(3\cos\theta_{1}-\cos\theta)^{3}}{\cos\theta}
\frac{Q^{2}}{\pi\sin\theta}\frac{d}{d\theta}
\left(\frac{\cos\theta-\cos\theta_{1}}{3\cos\theta_{1}-\cos\theta}
\right)d\theta-\frac{1}{a(4\pi
a^{3})^{2}\cos\theta_{1}}\\\nonumber&\times&\int_{0}
^{\theta_{1}}(3\cos\theta_{1}-\cos\theta)^{2}
\left(\frac{Q^{2}}{\sin\theta}\right)^{2}
d\theta+\frac{G\cos\theta_{1}}{4c^{4}}
\int_{0}^{\theta_{1}}(3\cos\theta_{1}-\cos\theta)^{2}\\\label{75}
&\times&\frac{Q_{0}^{2}}{\sin\theta\cos^{2}
\theta}d\theta-\frac{5Q^{2}}{8\pi
a^{3}}\int_{0}^{\theta_{1}}\frac{(3\cos\theta_{1}-\cos\theta)
^{2}}{\sin^{3}\theta}d\theta,
\end{eqnarray}
where $\theta_{1}=\sin^{-1}\left(\frac{R}{a}-\frac{b}{R}\right)$.
Solving the integrals and setting $\omega^2=0$, we obtain exact
condition for marginal stability. The values of $\Gamma_{c}$ for any
assigned value of $\theta$ are found such that $\gamma$ should be
less than certain $\Gamma_{c}$ for the existence of dynamical
instability. In Newtonian limit, $\Gamma$ takes finite values for
marginal stability, i.e., $\Gamma>\frac{4}{3}+\frac{8Q^2}{21}$. We
calculate the radii of marginal stability and $\Gamma$ for
homogeneous model of gaseous sphere by taking different values of
charge which show finite values of $\Gamma$. We observe that radius
$\frac{R}{R_{N}}\rightarrow\infty$ for $\Gamma<0$ which leads to the
expansion while $\frac{R}{R_{N}}$ remains positive for $\Gamma>0$
showing marginal stability of gaseous model. The corresponding
results are given in Table \textbf{1}.
\begin{table}[bht]
\textbf{Table 1:} \textbf{Adiabatic Index and Radii for Dynamical
Stability of Homogeneous Sphere} \vspace{0.5cm} \centering\\
\begin{small}
\begin{tabular}{|c|c|c|c|}
\hline\textbf{$\theta_{1}$}&\textbf{$R/R_{N}$}&
$\Gamma_{c}$ for $Q=0.2$&$\Gamma_{c}$ for $Q=0.6$\\
\hline{$0^\textmd{o}$}&{$\infty$}&{-0.0182}&{-0.1622}\\
\hline{$10^\textmd{o}$}&{33.163}&{0.1127}&{0.1776}\\
\hline{$20^\textmd{o}$}&{8.549}&{0.1275}&{0.2278}\\
\hline{$25^\textmd{o}$}&{5.598}&{0.1319}&{0.3192}\\
\hline{$30^\textmd{o}$}&{4.000}&{0.3766}&{1.2643}\\
\hline{$35^\textmd{o}$}&{3.0396}&{3246.43}&{1970.41}\\
\hline{$40^\textmd{o}$}&{2.4203}&{5918.49}&{2527.00}\\
\hline{$50^\textmd{o}$}&{1.704}&{6594.94}&{4352.86}\\
\hline{$60^\textmd{o}$}&{1.333}&{6822.02}&{5631.08}\\
\hline
\end{tabular}
\end{small}
\end{table}
\begin{figure}\center
\epsfig{file=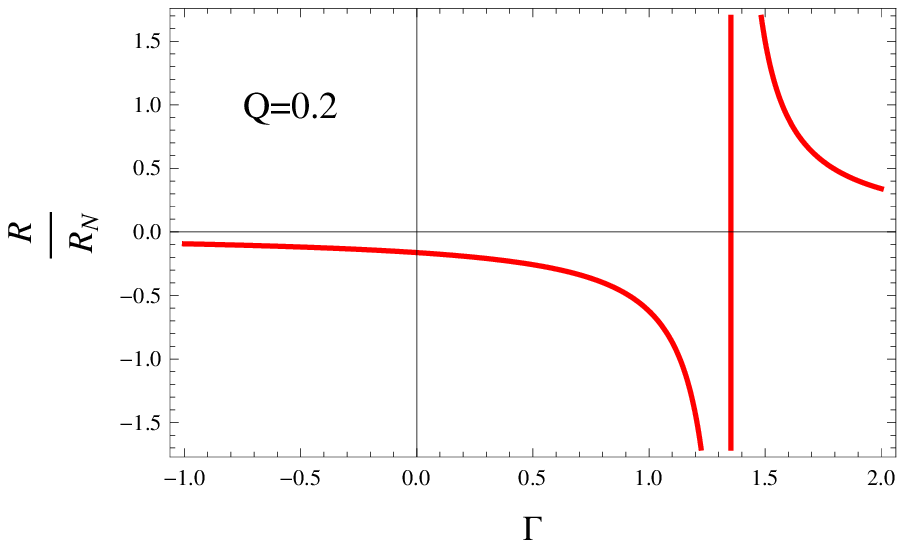,width=0.4\linewidth}\epsfig{file=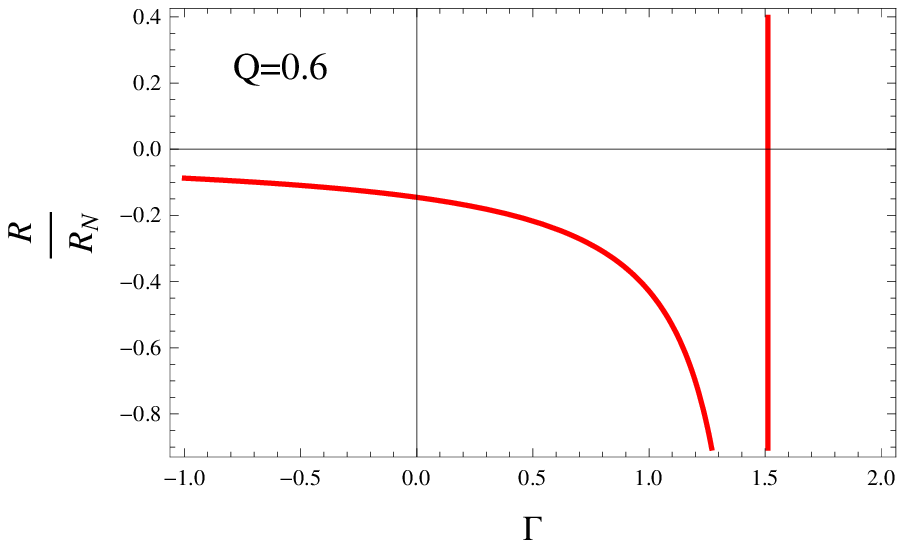,width=0.4\linewidth}\\
\epsfig{file=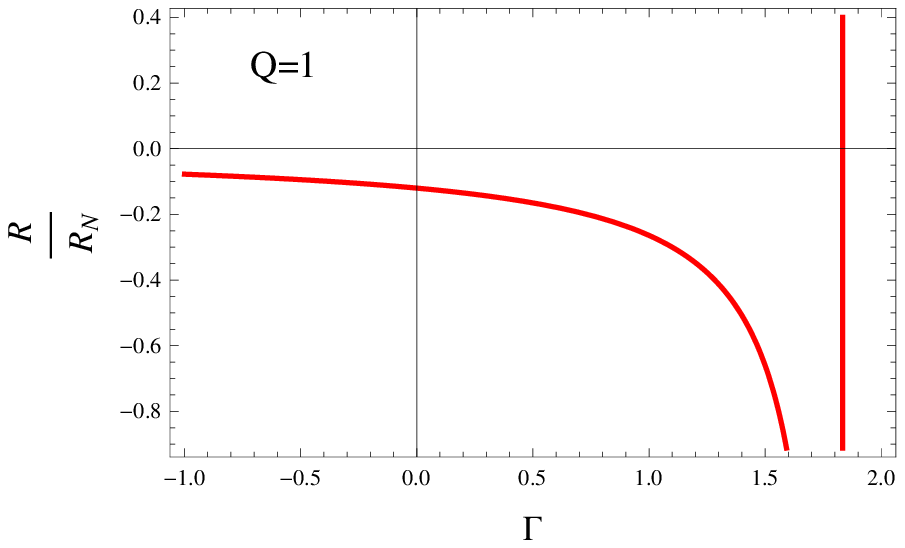,width=0.4\linewidth}\epsfig{file=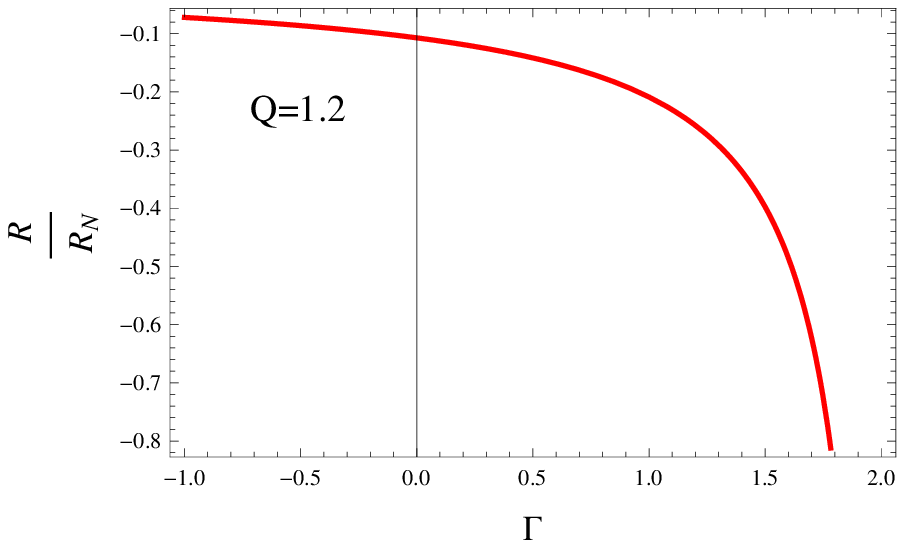,width=0.4\linewidth}\\
\caption{Plots of $\frac{R}{R_{N}}$ for dynamical
stability/instability of homogeneous sphere corresponding to
different values of Q.}
\end{figure}

When $\omega^2<0$, the perturbation diverges exponentially either by
expansion or contraction which yields stellar dynamical instability.
In the limit $\theta_{1}\rightarrow 0$, the condition for dynamical
instability is
\begin{equation}\label{78}
\Gamma-\frac{4}{21}\left(4Q^2+7\right)<\frac{14}{43}\theta_{1}^2
=\frac{14}{43}\left
[\frac{R^2}{a^2}-\frac{b^2}{R^2}\right].
\end{equation}
In terms of inertial mass, this takes the form
\begin{equation}\label{80}
R<\frac{14}{43\left[\Gamma-\frac{4}{21}\left(4Q^2+7\right)
\right]}\left[\frac{2GM}{c^2}-\frac{GQ^2}{Rc^4}\right],
\end{equation}
which can be written as
\begin{equation}\label{80}
\frac{R}{R_{N}}<\frac{K}{\left[\Gamma-\frac{4}{21}\left(4Q^2+7\right)
\right]},
\end{equation}
where $K=\frac{14}{43}$ for the homogeneous sphere. This means that
if $\Gamma$ exceeds $\frac{4}{3}+\frac{8Q^2}{21}$ by a small amount,
the dynamical instability can be prevented till the mass contracts
to the RN radius. If the radius of gaseous mass is greater than
$R_{N}$, it remains stable. The ranges for instability of
homogeneous spherical system are shown in Figure \textbf{1}. Since
the radius of stability is a factor of the RN radius, so the ratio
$\frac{R}{R_{N}}$ should be greater than or equal to zero for
physical results. We consider different values of charge and find
that $0.1<Q<0.4$ and $\Gamma>1.5$ provide valid radii ranges for the
stability of sphere. For $\Gamma<1.5$, we only have unstable radii
along with un-physical region corresponding to $Q>0.4$. Also, we
observe that the gaseous sphere becomes unstable forever with larger
values of charge, i.e., $Q>1.2$. It is obvious from the graph that
the radius of stability is greater than $R_{N}$ for
$\Gamma>\frac{4}{3}+\frac{8Q^2}{21}$.

\subsection{Relativistic Polytropic Model}

In relativistic polytropic models, pressure and energy density can
be expressed in terms of a single function $\Theta$ as \cite{7}
\begin{equation}\label{81}
p=p_{0}\Theta^{n+1}, \quad \sigma=\sigma_{0}\Theta^n,
\end{equation}
where $p_{0}$ and $\sigma_{0}$ represent respective values at center
and $n$ denotes the polytropic index. The polytropic models are the
generalized form of the classical Lane-Emden equation which can be
obtained from the equations of hydrostatic equilibrium. Let
\begin{equation}\label{82}
\xi=\frac{r}{a},
\end{equation}
where $a=\left(\frac{q(n+1)c^4}{4\pi
G\sigma_{0}}\right)^{\frac{1}{2}}$ and $q=
\frac{p_{0}}{\sigma_{0}}$. We can reduce Eqs.(\ref{23}) and
(\ref{24}) to the pair of equations  which express $\Theta$ as a
function of $\xi$
\begin{eqnarray}\nonumber
&&\left(\frac{\xi^2}{1+q\Theta}-\frac{c^2Q^2q(n+1)}{32\pi
G\sigma_{c}}\right)\left(1-\frac{2Vq(n+1)}{\xi}+\frac{Q^2Vq(n+1)}{\xi
M}\right)\frac{d\Theta}{d\xi}\\\label{82a}&&+V-\frac{Q^2V}{M}+
q\Theta\frac{dV}{d\xi}=0,
\end{eqnarray}
\begin{equation}\label{82b}
\frac{dV}{d\xi}=\xi\Theta^n.
\end{equation}
We assume pN approximation of the form
\begin{equation}\label{88}
\Theta=\theta+q\Phi,
\end{equation}
where $\Phi$ is an arbitrary function, $\theta$ represents classical
Lane-Emden function and $q$ is treated as a small constant. Using
Eqs.(\ref{82a}) and (\ref{82b}), the classical Lane-Emden equation
becomes
\begin{equation}\label{82c}
\frac{d^2\theta}{d\xi^2}+\frac{2}{\xi}\frac{d\theta}{d\xi}
+\left(1-\frac{Q^2}{M}\right)\theta^n=\frac{c^2Q^2}{32\pi
G\sigma_{c}}.
\end{equation}
Equation (\ref{61}) in terms of $\Theta$ and $\xi$ takes the form
\begin{eqnarray}\nonumber
&&\frac{(a\omega)^{2}}{q}\int^{\xi_{1}}_{0}\Theta^n(1+q\Theta)
\xi^2\eta^2e^{(3\lambda-\nu)/2}d\xi=2(n+1)q\int^{\xi_{1}}_{0}
\Theta^{(2n+1)}(1+q\Theta)\\\nonumber&&\times\xi^2\eta^2
e^{(3\lambda+\nu)/2}d\xi+4(n+1)\int^{\xi_{1}}_{0}\Theta^n
\xi\eta^2e^{(\lambda+\nu)/2}\left(1-\frac{q\xi(n+1)}
{4(1+q\Theta)}\right)d\xi\\\nonumber&&+\Gamma
\int^{\xi_{1}}_{0}\frac{\Theta^{n+1}}{\xi^2}e^{(\lambda+3\nu)/2}
\left(\frac{d}{d\xi}\left[\eta\xi^2e^{-\nu/2}\right]\right)^2
d\xi+\left(\frac{1}{2\pi
a^3}\right)^2\int^{\xi_{1}}_{0}\frac{e^{(\lambda+\nu)/2}}
{\Theta^n(1+q\Theta)}\\\nonumber&&\times\left(\frac{\eta
Q}{\xi^4}\right)^2\left[\frac{\xi}{2a}-Q^2\right]d\xi-\frac{1}{2\pi
a^3}\int^{\xi_{1}}_{0}\frac{\eta^2}{\xi^3}e^{(\lambda+\nu)/2}
\left[1+\frac{q(n+1)Q^2}{1+q\Theta}\frac{d\Theta}{d\xi}\right.
\\\nonumber&&\left.-\frac{GQ_{0}^2}{\xi^4}\right]d\xi+\frac{1}
{4\pi}\int^{\xi_{1}}_{0}\frac{\eta^2}{1+q\Theta}
e^{(\lambda+\nu)/2}\frac{dQ^2}{d\xi}\left[q\frac{d\Theta}
{d\xi}-\frac{1}{\Theta^n\left(a^3\xi^3\right)\frac{dQ^2}
{d\xi}}\right]d\xi\\\nonumber&&-\frac{1}{8\pi
a^3}\int^{\xi_{1}}_{0}\frac{\eta}{\xi^2}e^{(\lambda+\nu)}
\frac{d^2}{dr^2}\left(Q^2\right)d\xi+ \int^{\xi_{1}}_{0}\xi^2\eta
e^{(\nu-\lambda)/2}\frac{d}{d\xi}\left[\frac{\beta
Q_{0}\widetilde{Q}e^{\lambda_{0}}}{a\xi^3}\right]d\xi
\\\label{83}&&+\frac{1}{4\pi a^2}\int^{\xi_{1}}_{0}\xi^2\eta
e^{(\nu+\lambda)/2}\frac{d}{d\xi}\left(\frac{Q_{0}\delta
Qe^{-\nu/2}}{\xi^4}\right)d\xi.
\end{eqnarray}

The pN approximation treats the effects of GR as first order
corrections. We can write
\begin{eqnarray}\label{86}
\Gamma_{c}-\frac{4}{21}(4Q^2+7)=Cq, \\\label{87}
R_{c}=\frac{K}{\Gamma_{c}-\frac{4}{21}(4Q^2+7)}
\left[\frac{2GM}{c^2}-\frac{GQ^2}{Rc^4}\right],
\end{eqnarray}
where $C$ and $K$ are constants depending on density distribution.
The pN approximation yields
\begin{equation}\label{88a}
e^{-\lambda}=1+2q(1+n)\xi\frac{d\theta}{d\xi}-
\frac{GQ^2(1+n)q}{M}\frac{d\theta}{d\xi},
\end{equation}
\begin{equation}\label{88b}
e^{\nu}=1-2q(1+n)[\theta+\xi_{1}|\tilde{\theta}_{1}|]
+\frac{Q^2(1+n)q|\tilde{\theta}_{1}|}{M},
\end{equation}
where
$|\tilde{\theta}_{1}|=-\left(d\theta/d\xi\right)_{\xi=\xi_{1}}$.
Using the relations of $p_{c}$ and $\sigma_{c}$ for polytropes in
terms of $M$, $Q$ and $R$, it follows that
\begin{equation}\label{89}
q=\frac{1}{2(n+1)\xi_{1}|\tilde{\theta}_{1}|}
\left[\frac{2GM}{Rc^2}-\frac{GQ^2}{R^2c^4}\right].
\end{equation}
We calculate $\Gamma$ with different values of $Q$ for the emergence
of dynamical instability. The numerical values of $\Gamma$ for the
polytropes of index 3 are given in Table \textbf{2}. Similarly, the
constants $C$ and $K$ for polytropes are given by the relation
\begin{equation}\label{89}
K=\frac{C}{2(n+1)\xi_{1}|\tilde{\theta}_{1}|}.
\end{equation}
\begin{table}
\textbf{Table 2:} \textbf{Adiabatic Index with Different Values of
Charge for Dynamical Instability of Polytropes with Index 3} \vspace{0.5cm}\centering\\
\begin{small}
\begin{tabular}{|c|c|c|c|}
\hline\textbf{$q$}&{$\Gamma_{c}$ for $Q=0.2$}&
$\Gamma_{c}$ for $Q=0.4$&$\Gamma_{c}$ for $Q=0.6$\\
\hline{$0.015$}&{$1.3500$}&{1.3593}&{1.4715}\\
\hline{$0.040$}&{1.3527}&{1.3983}&{1.4744}\\
\hline{$0.1$}&{1.3586}&{1.4043}&{1.4805}\\
\hline{$0.2$}&{1.3686}&{1.4143}&{1.4905}\\
\hline{$0.5$}&{1.3953}&{1.4440}&{1.5204}\\
\hline
\end{tabular}
\end{small}
\end{table}
In order to determine the radii from Eq.(\ref{87}), we need to
calculate $K$ whose value depends upon the polytropic index,
Lane-Emden function and charge. Different polytropic indices lead to
different stellar structures such that the configurations with $n<5$
are considered to be realistic stars \cite{11}. For $n=0$, we solve
the Lane-Emden equation analytically corresponding to different
values of $Q$ and find the values of $\theta$ but we solve this
equation numerically for $n=2,3,4$ as shown in Figures \textbf{2-3}.
The values of constants $C$ and $K$ for $n=0$ are given in Table
\textbf{3}. We see that the values of $K$ decrease gradually by
increasing the value of charge.
\begin{table}
\textbf{Table 3:} \textbf{Values of Constants $C$
and $K$ with Different Values of Charge for Polytropes of
Index $0$} \vspace{0.5cm}\centering\\
\begin{small}
\begin{tabular}{|c|c|c|c|}
\hline\textbf{$C$}&{$K$ for $Q=0.2$}&
$K$ for $Q=0.4$&$K$ for $Q=0.6$\\
\hline{$0.243$}&{0.6458}&{0.6372}&{0.576}\\
\hline{$0.826$}&{$2.1953$}&{2.1662}&{1.9608}\\
\hline{$1.205$}&{3.2034}&{3.161}&{2.8612}\\
\hline{$1.8095$}&{4.809}&{4.745}&{4.296}\\
\hline
\end{tabular}
\end{small}
\end{table}
\begin{figure}\center
\epsfig{file=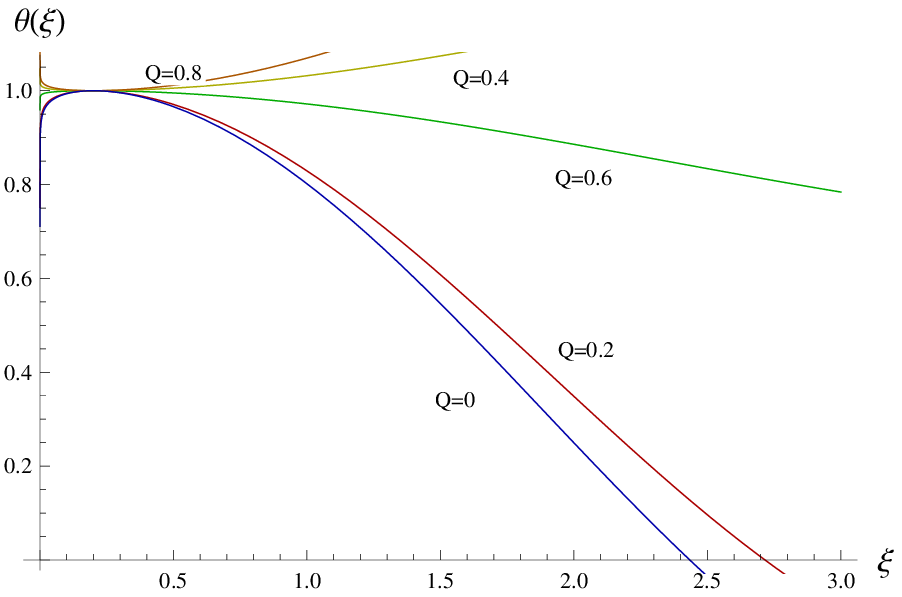,width=0.45\linewidth}\epsfig{file=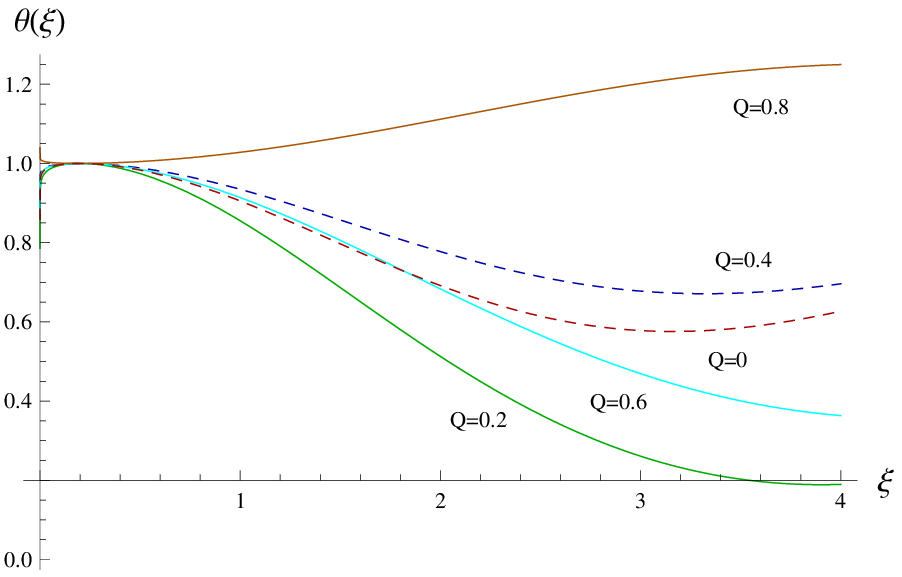,
width=0.47\linewidth}\\\caption{Plots of classical Lane-Emden
function for polytropes of index $n=1,2$ corresponding to different
values of charge.}
\end{figure}
\begin{figure}\center
\epsfig{file=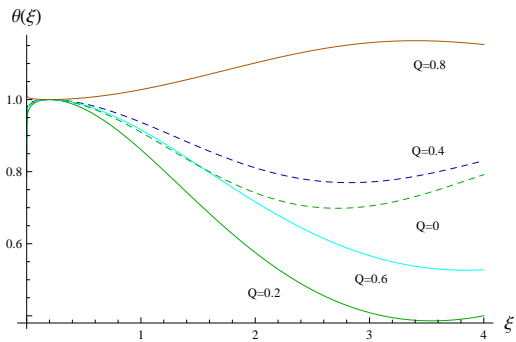,width=0.5\linewidth}\\
\caption{Plots of classical Lane-Emden function for $n=3$
corresponding to different values of charge.}
\end{figure}

Inserting the values of $e^{\nu}$ and $e^{-\lambda}$ in
Eq.(\ref{83}) and neglecting second as well as higher order terms in
$q$, we obtain
\begin{eqnarray}\nonumber
&&\frac{(a\omega)^{2}}{q}\left\{\int^{\xi_{1}}_{0}
\theta^n\xi^2\eta^2d\xi+\int^{\xi_{1}}_{0}H(\xi)\theta^{n-1}
\xi^2\eta^2d\xi\right\}=\Gamma q(n+1)\int^{\xi_{1}}_{0}
\frac{\theta^{n+1}}{\xi^2}\\\nonumber&&\times\left\{\left(\frac{GQ^2}
{2M}-\xi\right)+3|\tilde{\theta}_{1}|
\left(\frac{Q^2}{2M}-\xi_{1}\right)-3\theta\right\}\left
[\frac{d}{d\xi}(\xi^2\eta)\right]^2d\xi
+\Gamma\int^{\xi_{1}}_{0}H(\xi)\\\nonumber&&\times
\frac{\theta^{n+1}}{\xi^2}\left\{\frac{d}{d\xi}
\left(\xi^2\eta\right)\right\}^2d\xi+q(n+1)\int^{\xi_{1}}_{0}
I(\xi)\theta^{n-1}\xi\eta^2d\xi+q(n+1)
\int^{\xi_{1}}_{0}\frac{d\theta}{d\xi}\\\nonumber&&\times
\left[\left(\frac{GQ^2}{2M}-\xi\right)+|\tilde{\theta}_{1}|
\left(\frac{Q^2}{2M}-\xi_{1}\right)-\theta\right]Y(\xi)d\xi
+q(n+1)\int^{\xi_{1}}_{0}\theta^{n+1}\xi^2\eta^2
\\\nonumber&&\times Z(\xi)d\xi+\frac{1}{(2\pi)^2}
\int^{\xi_{1}}_{0}\frac{\eta^2Q^2}{(a^3\xi^2)^3}
\left[\frac{\xi}{2a}-Q^2\right]d\xi-\frac{1}{8\pi}
\int^{\xi_{1}}_{0}\frac{\eta}{a^3\xi^2}\frac{d^2}{d\xi^2}
(Q^2)d\xi\\\label{91}&&-\frac{1}{2\pi}\int^{\xi_{1}}_{0}
\frac{GQ^2}{a^3\xi^4}d\xi+\int^{\xi_{1}}_{0}\xi^2\eta
\frac{d}{d\xi}\left(\frac{\beta
Q_{0}\widetilde{Q}}{a\xi^3}\right)d\xi,
\end{eqnarray}
where
\begin{eqnarray}\nonumber
H(\xi)&=&2\theta^2(n+2)+3(n+1)\left(\frac{GQ^2}{2M}-\xi\right)
\frac{d\theta}{d\xi}+
\left(2\xi_{1}-\frac{Q^2}{2M}\right)\\\label{92}&\times&
(n+1)|\tilde{\theta}_{1}|,\\\label{93}
I(\xi)&=&-4\theta(n+1)\frac{d\theta}{d\xi}\left[-\xi
\frac{d\theta}{d\xi}-\theta-\xi|\tilde{\theta}_{1}|+
\frac{Q^2}{2M}\left(G\frac{d\theta}{d\xi}+
|\tilde{\theta}_{1}|\right)\right],\\\nonumber
Y(\xi)&=&\frac{\eta^2Q^2}{2\pi^2(a^3\xi^2)^2}\left(\frac{\xi}
{2a}-Q^2\right)-\frac{\eta^2}{2\pi a^3\xi^3}+\frac{GQ_{0}^2}{2\pi
a^3\xi^4}-\frac{\eta}{4\pi
a^3\xi^2}\\\label{94}&\times&\frac{d^2}{d\xi^2}(Q^2)+
\xi^2\eta\frac{d}{d\xi}\left\{\frac{\beta
Q_{0}\widetilde{Q}}{a\xi^3}+\eta\xi^2\frac{d}{d\xi}
\left(\frac{Q_{0}\delta Q}{\xi^4}\right)\right\}, \\\label{95}
Z(\xi)&=&3\theta\left(\frac{Q^2}{2M}-\xi\right)\frac{d\theta}{d\xi}
+\theta|\tilde{\theta}_{1}|\left(\frac{Q^2}{2M}-2\xi_{1}\right)-
2\theta^2.
\end{eqnarray}
In pN approximation, we are interested to find the condition for
marginal stability of polytropic configuration by taking $\eta=\xi$
and $\omega^2=0$ so that Eq.(\ref{91}) takes the form
\begin{figure}\center
\epsfig{file=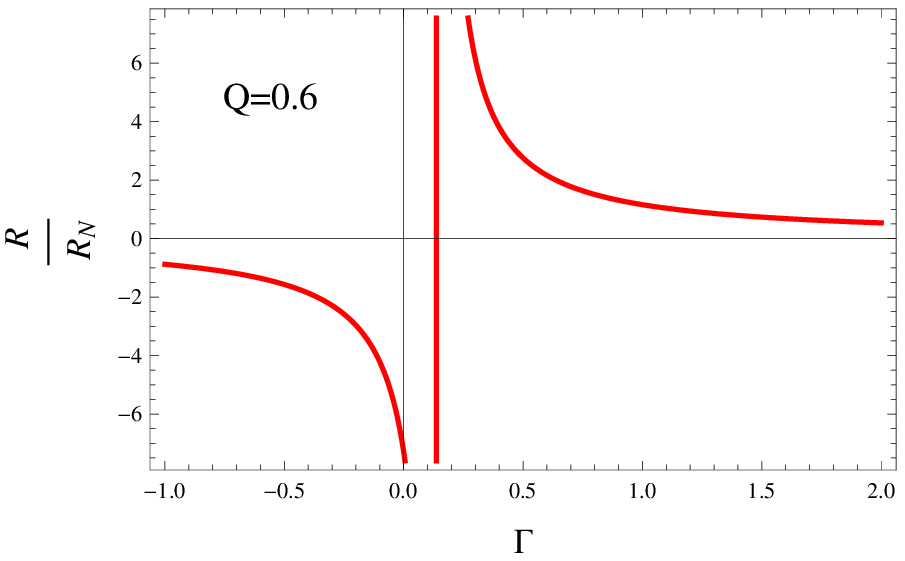,width=0.4\linewidth}\epsfig{file=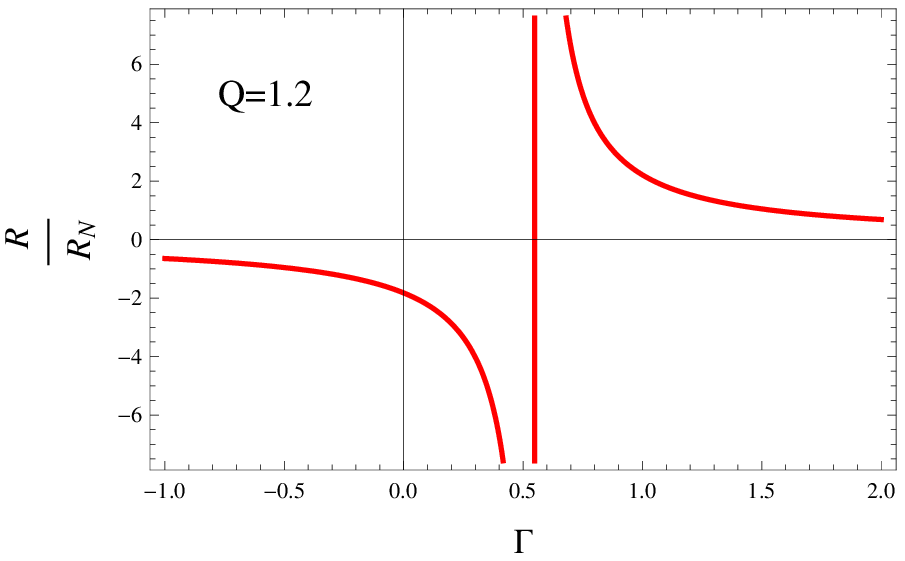,width=0.4\linewidth}\\
\epsfig{file=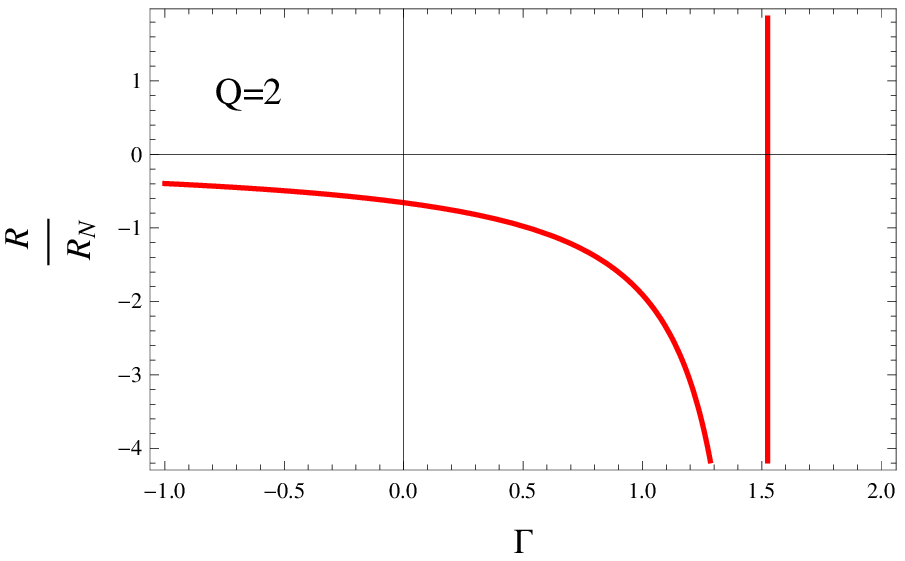,width=0.4\linewidth}\epsfig{file=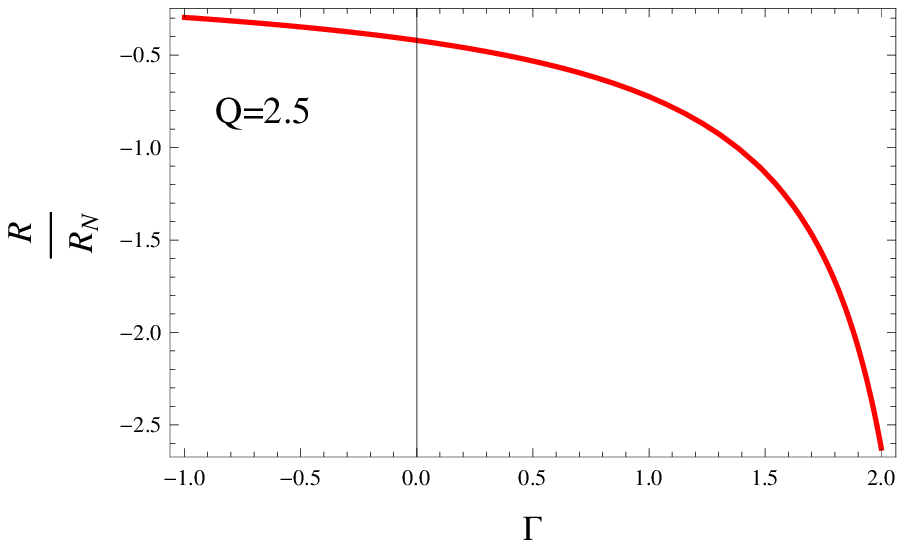,width=0.4\linewidth}\\
\caption{Plots for radii of stability/instability corresponding to
$n=1$ and different values of charge.}
\end{figure}
\begin{eqnarray}\nonumber
&&9\left(\Gamma-\frac{4}{3}-\frac{8Q^2}{21}\right)\int^{\xi_{1}}_{0}
\theta^{n+1}\xi^2d\xi+q(n+1)
\left[3\int^{\xi_{1}}_{0}\theta^nY(\xi)\widetilde{Y}(\xi)d\xi\right.
\\\nonumber&&\left.+ \int^{\xi_{1}}_{0}\theta^{n-1}I(\xi)\xi^3d\xi+
\int^{\xi_{1}}_{0}\theta^{n+1}\xi^4Z(\xi)d\xi\right]+\frac{1}{8\pi}
\int^{\xi_{1}}_{0}\frac{\widetilde{Z}(\xi)}{a^3\xi^3}d\xi\\\nonumber
&&+ \int^{\xi_{1}}_{0}\xi^3\frac{d}{d\xi}\left(\frac{\beta
Q\widetilde{Q}}{a\xi^3}\right)d\xi=0,\\\label{94}
\end{eqnarray}
with
\begin{figure}\center
\epsfig{file=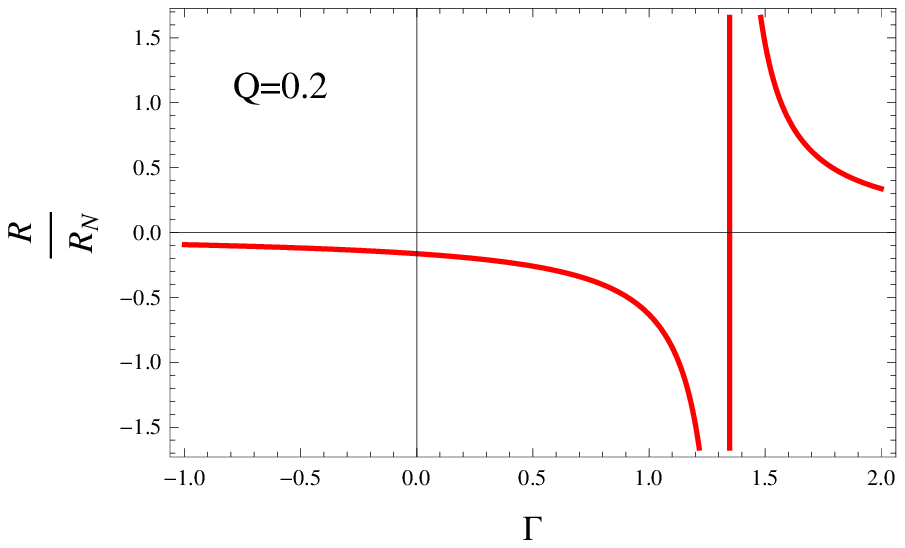,width=0.4\linewidth}\epsfig{file=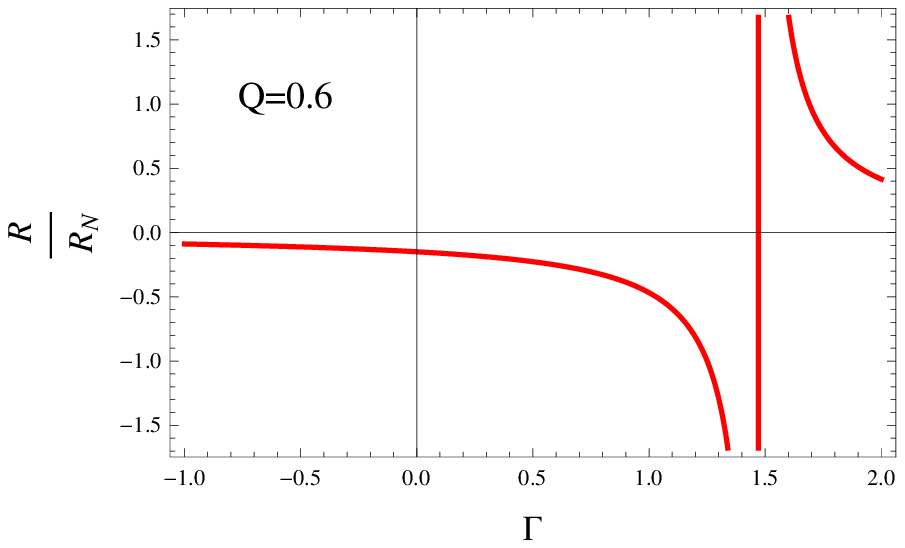,width=0.4\linewidth}\\
\epsfig{file=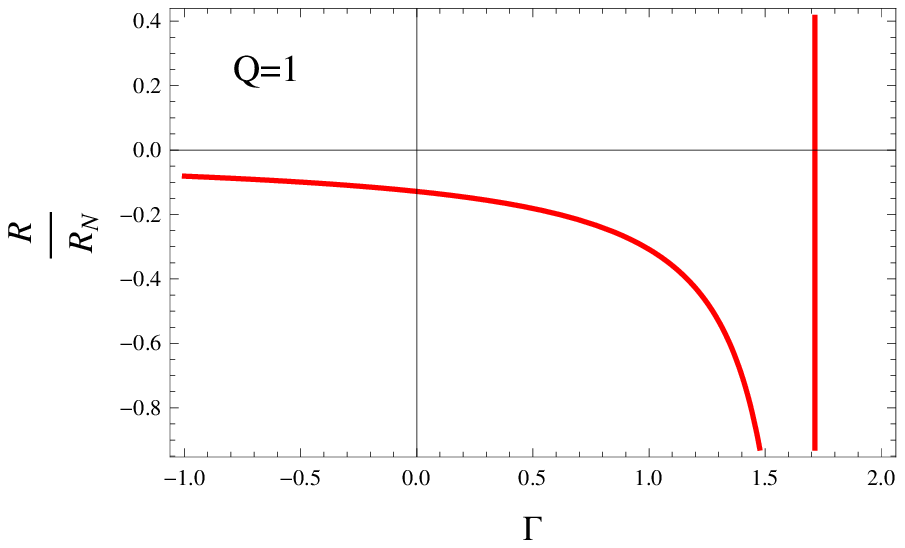,width=0.4\linewidth}\epsfig{file=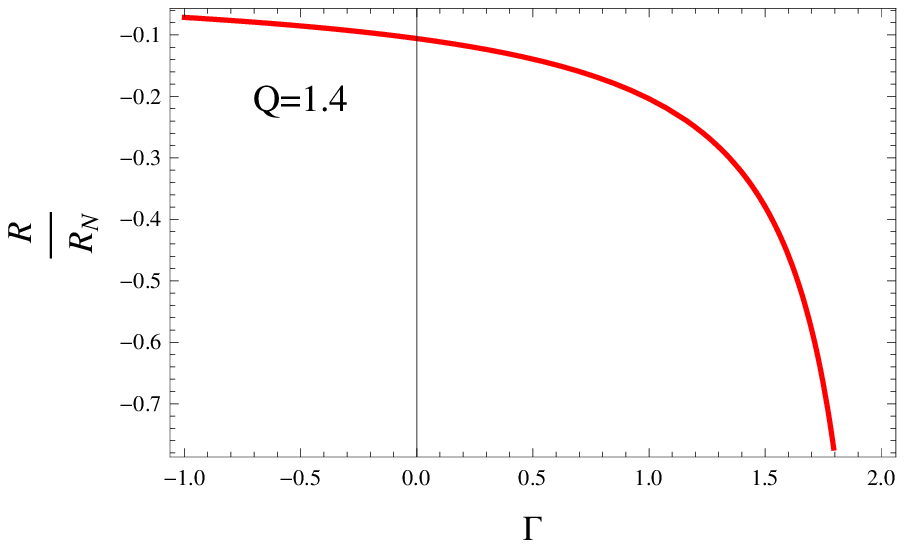,width=0.4\linewidth}\\
\caption{Plots for radii of stability/instability corresponding to
$n=2$ and different values of charge.}
\end{figure}
\begin{eqnarray}\nonumber
\widetilde{Y}(\xi)&=&\left(3\Gamma\xi^2\theta+\theta^{-n}\right)
\left\{\frac{d\theta}{d\xi}\left(\frac{GQ^2}{2M}-\xi\right)
+3|\tilde{\theta}_{1}|\left(\frac{Q^2}{2M}-\xi_{1}\right)-3\theta
\right\},\\\label{95}\\\label{95a}
\widetilde{Z}(\xi)&=&4Q^2\left(G+\frac{\xi}{2M}-Q^2\right)-
\xi^3\left(1+\frac{d^2}{d\xi^2}(Q^2)\right).
\end{eqnarray}
\begin{figure}\center
\epsfig{file=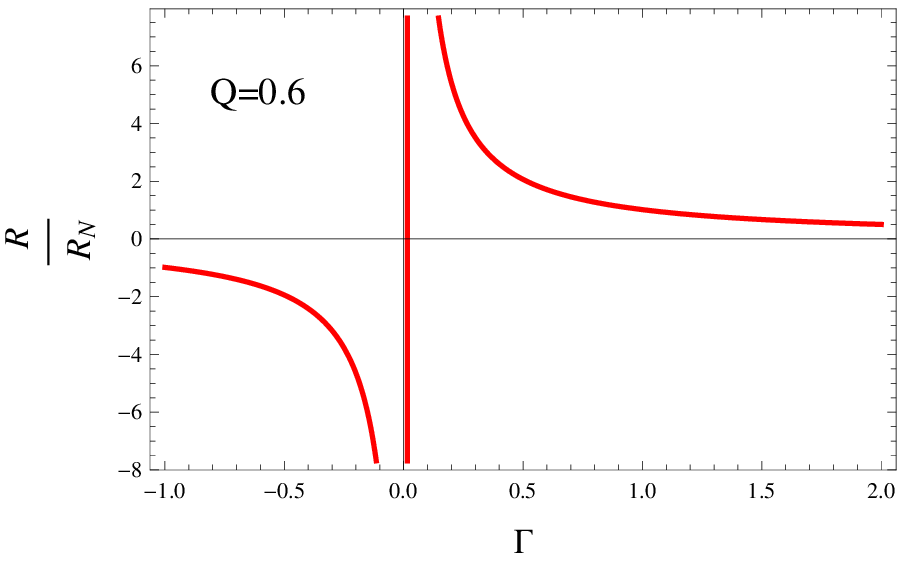,width=0.4\linewidth}\epsfig{file=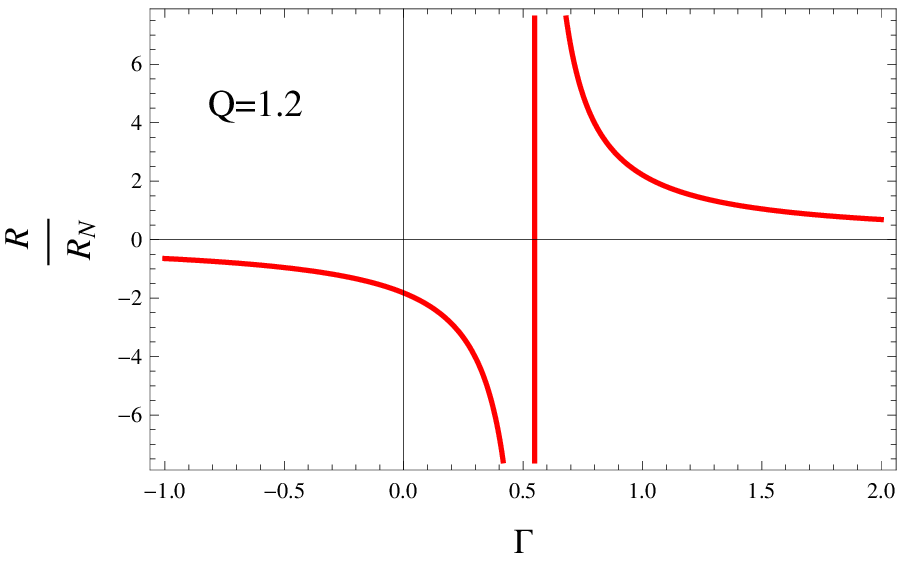,width=0.4\linewidth}\\
\epsfig{file=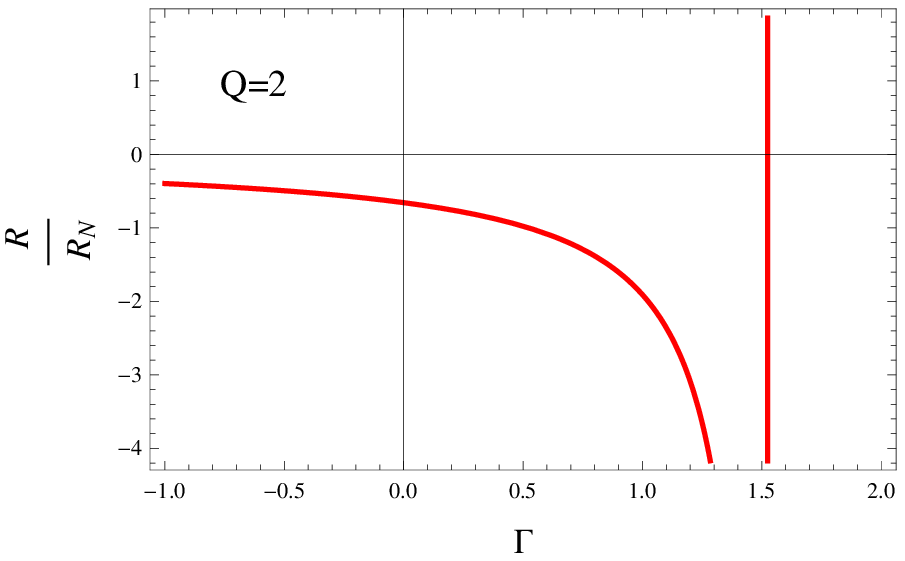,width=0.4\linewidth}\epsfig{file=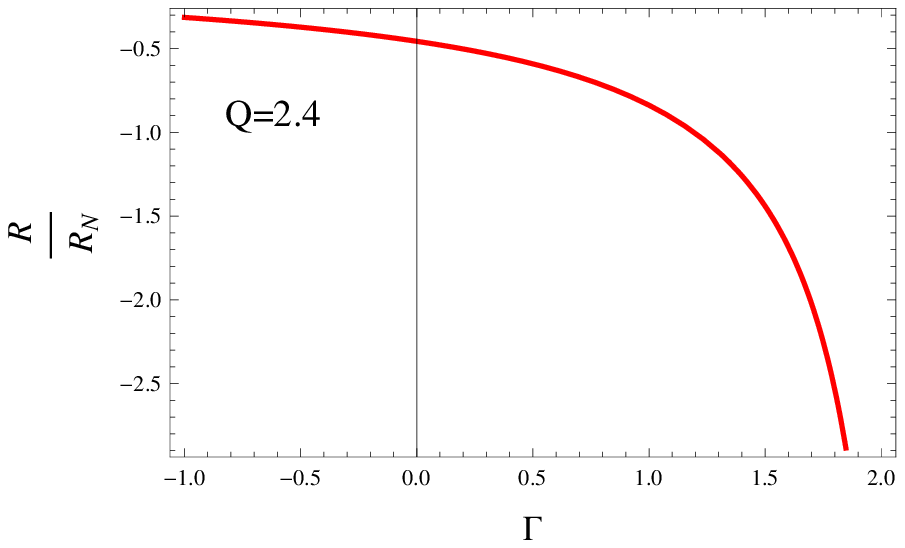,width=0.4\linewidth}\\
\caption{Plots for radii of stability/instability corresponding to
$n=3$ and different values of charge.}
\end{figure}
In pN limit, the dynamical stability will require that
$\Gamma>\Gamma_{c}=\frac{4}{3}+\frac{8Q^2}{21}+\epsilon$, where
$\epsilon$ is a small quantity depending on $q$. To check the
conditions for marginal stability as well as dynamical instability,
we calculate values of $K$ and plot different radii corresponding to
polytropic indices $n=1,2,3$ as shown in Figures \textbf{4-6}. It is
found that $K$ attains negative values for $n=1,2,3$, so we take
both positive and negative values of $\Gamma$ to obtain physically
viable values of radii. Figures \textbf{4} and \textbf{5} show
viable radii for $\Gamma>\frac{4}{3}+\frac{8Q^2}{21}$ corresponding
to $n=1,2$. For $n=1$, we find that radii of stability along with
non-physical region appear for $Q=0.6$ and positive values of
$\Gamma$ while negative values of $\Gamma$ show the emergence of
instability. The region of instability gets larger by increasing $Q$
for both positive and negative values of $\Gamma$ and the radii of
marginal stability vanishes for $Q=2$. We observe that the the
corresponding polytropic model becomes unstable for $Q>2.4$ (Figure
\textbf{4}).

For $n=2$, we analyze stable radii for $1.5<\Gamma<2$ and unstable
radii for small values of $\Gamma$ (both positive and negative)
corresponding to $Q<1$. The stability radius tends to decrease which
leads to unstable region for $Q=1$. We find that the non-physical
region disappears and the polytropic model will remain unstable
forever with $Q\geq1.4$ (Figure \textbf{5}). For $n=3$, we have
viable ranges for radii with $0<\Gamma<2$, i.e., $\Gamma$ can be
less than $\frac{4}{3}+\frac{8Q^2}{21}$ for stable stellar
structures (Figure \textbf{6}). We find both stable and unstable
radii for polytropic model with $0.1<Q<2$ which changes to unstable
radii for $Q\geq2$. The dynamical instability occurs when gaseous
mass contracts to $R_{N}$.

\section{Conclusions}

This paper is devoted to investigate the role of electric charge on
dynamical stability of spherical gaseous masses under radial
oscillations. We have perturbed the system using Eulerian and
Lagrangian radial perturbations to obtain linearized dynamical
equations as well as perturbed pressure. This perturbed pressure in
terms of adiabatic index is found using conservation of baryon
number. The variational principle has been applied to the perturbed
dynamical equations to formulate conditions of dynamical
instability. We have also discussed conditions for dynamical
instability of homogeneous sphere and relativistic polytropes in
Newtonian as well as pN regimes.

We have found the values of adiabatic index $\Gamma$ as well as
radii for marginal stability of homogeneous sphere (Table
\textbf{1}). It turns out that $\Gamma$ takes finite positive values
less or greater than $\frac{4}{3}+\frac{8Q^2}{21}$ corresponding to
different values of charge at Newtonian limit. The radius
$\frac{R}{R_{N}}$ approaches to infinity for $\Gamma<0$ which leads
to expansion rather than collapse. In pN limit, the dynamical
instability occurs if $\Gamma$ exceeds $\frac{4}{3}+\frac{8Q^2}{21}$
by a small quantity and the gaseous mass is contracted to RN radius.
We have found that $0.1<Q<0.4$ and $\Gamma>1.5$ provide valid radii
ranges for stability of sphere. It turns out that only unstable
radii exist corresponding to $Q>1.2$.

We have also discussed the stability/instability conditions for
relativistic polytropic models of indices $0,1,2$ and $3$. For
$n=3$, we have evaluated different values of $\Gamma$ for dynamical
instability of polytropes and found that
$\Gamma>\frac{4}{3}+\frac{8Q^2}{21}$ (Table \textbf{2}). In order to
discuss realistic models, we have evaluated radius of instability
for different polytropic structures. We have also calculated the
values of $C$ and $K$ for $n=0$ (Table \textbf{3}), which show that
$K$ decreases gradually by increasing the values of charge. For
$n=1,2$, we have viable radii corresponding to
$\Gamma>\frac{4}{3}+\frac{8Q^2}{21}$. For $n=1$, we have found that
radii of stability along with non-physical region exist for $Q=0.6$
and $\Gamma>0$ while negative values of $\Gamma$ show the emergence
of instability. The region of instability increases by increasing
$Q$ for both positive and negative values of $\Gamma$ and we find
only unstable radii for $Q>2.4$. For $n=2$, we have found both
stable and unstable radii for $1.5<\Gamma<2$. The stability radius
tends to decrease gradually which leads to unstable region for
$Q=1.4$. It is seen that $\Gamma$ can be less than
$\frac{4}{3}+\frac{8Q^2}{21}$ for $n=3$. We have analyzed both
stable and unstable radii for polytropic model with $0.1<Q<2$ which
changes to unstable radii for $Q\geq2$.

It is found that the dynamical instability occurs when the mass of
polytropic configuration approaches to the RN radius limit. We
observe that inclusion of charge in the gaseous sphere has
significant effects as compared to the analysis \cite{b}. For
charged homogeneous sphere, the system becomes stable for both
negative as well a as positive values of adiabatic index, while it
remains stable for $\Gamma>\frac{4}{3}$ without charge system
\cite{b}. For the charged polytropes with $n=1,2,3$, $\Gamma$ can
take both positive as well as negative values while $K$ becomes
negative. We also see that the radius of instability of polytropes
($n=3$) for RN case is greater than that of the Schwarzschild limit
showing that RN polytropes for $n=3$ are more stable. Finally, we
conclude that the presence of charge has substantial role in the
emergence of instability of gaseous sphere.

\end{document}